\documentclass[
  aip,
  jcp,
  longbibliography,
  amsmath,
  amssymb,
  reprint,
  numerical
]{revtex4-1}

\usepackage{graphicx}
\usepackage{dcolumn}
\usepackage{bm}
\usepackage[utf8]{inputenc}
\usepackage[T1]{fontenc}
\usepackage{mathptmx}
\usepackage{etoolbox}
\usepackage{placeins}
\usepackage{xcolor}
\usepackage[version=4]{mhchem}
\usepackage[hidelinks]{hyperref}

\usepackage{multirow}
\usepackage{siunitx}

\definecolor{draftgray}{gray}{0.45}

\makeatletter
\AtBeginDocument{%
\let\@affil@present@script\@affil@present@group
}
\def\@email#1#2{%
  \endgroup
  \patchcmd{\titleblock@produce}
    {\frontmatter@RRAPformat}
    {\frontmatter@RRAPformat
      {\produce@RRAP{*#1%
  \href{mailto:mchow@udel.edu}{mchow@udel.edu};
  \href{mailto:#2}{#2}}}
      \frontmatter@RRAPformat}
    {}{}
}%
\makeatother

\begin{document}

\preprint{AIP/PLACEHOLDER}

\title{Can AI Agents Reliably Optimize Computational Chemistry Codes?}

\author{Mathew Chow\textsuperscript{*}}
\affiliation{Department of Physics and Astronomy, University of Delaware, Newark, DE 19716, United States}

\author{Tao E. Li\textsuperscript{*}}
\email{taoeli@udel.edu}
\affiliation{Department of Physics and Astronomy, University of Delaware, Newark, DE 19716, United States}

\date[]{}

\begin{abstract}
Artificial intelligence (AI) approaches have opened new possibilities in computational and theoretical chemistry. AI agents can now execute computational workflows autonomously, and their use in scientific software development represents another promising direction. Herein, we present a general strategy for the controlled optimization of scientific software using a dual-agent harness. By using separate worker and controller agents for optimization and evaluation, respectively, we find that this approach can reduce computational costs while preserving numerical accuracy for the computational chemistry methods examined. We apply this strategy to TIP4P water simulations in LAMMPS, as well as to Coulomb and exchange matrix construction and linear-response time-dependent density functional theory (TDDFT) calculations in PySCF. Among these applications, the largest speedups are obtained for TDDFT. We then extend this strategy from optimizing established packages to developing efficient parallel CPU- and GPU-accelerated implementations of an initially serial molecular dynamics code, resulting in speedups of two to three orders of magnitude. Finally, we apply the protocol to open-ended algorithm development by constructing an adaptive self-consistent field approach that coordinates the use of multiple solvers. We find that the resulting approach can converge some challenging systems and occasionally reach lower-energy solutions than those obtained with the standalone solvers tested. These applications demonstrate that a strategy developed for improving code efficiency may also support the systematic development of computational methods. Overall, our benchmark study provides an example of AI-assisted advances in computational chemistry.
\end{abstract}

\maketitle
\thispagestyle{plain}
\pagestyle{plain}

\section{Introduction}
\label{sec:introduction}

Artificial intelligence (AI) and machine learning technologies are increasingly used in research across scientific computing.\cite{Ju2021, Jablonka2023, Cui2025} AI tools can help automate workflows and explore new approaches to scientific problems.\cite{Gadde2025, Tran2026} As commercially available large language models (LLMs) rapidly evolve and become more widely integrated into everyday research practice, there is also an interest in connecting these models to scientific software through specialized agent harnesses.\cite{Zou2025,Huang2025,Meng2026,Lee2026} These harnesses provide an infrastructure that enables agents to execute multi-step workflows, and this can inherently make computer simulations, and even some experiments,\cite{Bauer2026} more accessible especially to non-experts. AI agents can also facilitate scientific research in other ways. Recent studies in computer science have used LLM-based code generation together with repeated evaluation to improve general code performance.\cite{Romera-Paredes2024,Novikov2025} In chemistry and materials science, agents have been used to write code, perform simulations, analyze data,\cite{Meng2026,Shen2026} and even propose a new exchange-correlation functional.\cite{Duston2026} These examples showcase ways that AI can be used to solve increasingly difficult problems in science.

As AI is capable of writing long and complicated codes very quickly, it perhaps becomes increasingly intractable for researchers to review code at a granular level.
In this current paradigm shift, it is important to introduce robust and deterministic strategies to conduct AI-assisted scientific code implementation. Without this effort, human scientists may gradually lose control of AI-generated scientific code.

In this work, we devise a protocol that enables LLM-based AI agents to systematically and iteratively optimize computational chemistry code in a controllable manner. We show that this same strategy can be extended from improving computational efficiency to developing new numerical approaches. We first use this protocol to accelerate calculations in open-source classical molecular dynamics (MD) and quantum chemistry software packages. We then examine whether an initial serial MD implementation can be developed into efficient, parallelized CPU and GPU codes. Finally, we adapt this protocol to develop an approach that coordinates and switches between multiple self-consistent field (SCF) solvers. This progression enhances our understanding of AI, and the knowledge gained can help future advancements in computational and theoretical chemistry. 

An outline of this paper is as follows: In Section~\ref{sec:methods}, we discuss this AI-assisted optimization strategy and share a few of the decisions we made during the preparation of the optimization calculations. We then briefly discuss the relevant scientific methodologies underlying the codes that are being targeted for optimization. Section~\ref{sec:computational_details} provides computational details for the calculations performed in Section~\ref{sec:results_and_discussion}. Finally, we provide our concluding thoughts in Section \ref{sec:conclusions}. 

\section{Methods}
\label{sec:methods}

In this section, we introduce the structure of the dual-agent harness for scientific code optimization, followed by a brief overview of the involved computational chemistry approaches.

\subsection{Optimization Protocol}
\begin{figure}[!t]
  \centering
  \includegraphics[width=\columnwidth]{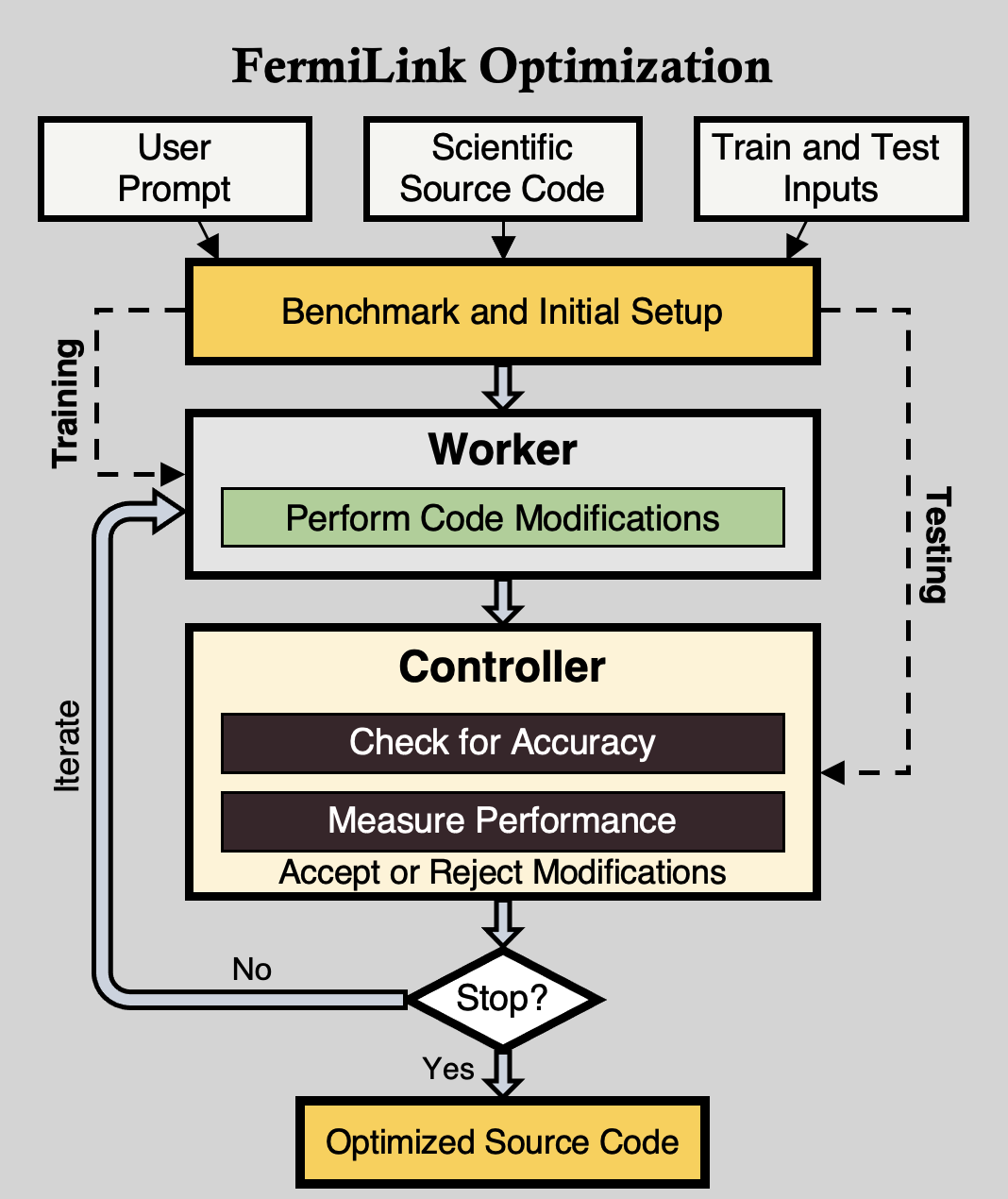}
 \caption{
  Schematic of the \texttt{optimize} workflow in the \textsc{FermiLink} agent harness. According to the supplied prompt file, a worker agent uses the training cases to propose targeted source code changes on the  scientific software package. A controller agent evaluates the code for computational performance and scientific accuracy with a different set of testing cases. Accepted changes are then retained for subsequent iterations, and the iterative procedure proceeds until a stopping threshold is met or a consecutive rejection streak is reached.
  }
  \label{fig:fermilink-optimize-scheme}
\end{figure}

The central idea of our optimization protocol is to use a feedback loop that enables successive revisions of scientific code between a worker agent and a controller agent. This strategy borrows training and testing design principles from neural network fitting\cite{LeCun2015,Butler2018,Mater2019,Unke2021} but focuses on changing the code instead of tuning model parameters or adjusting model weights. We implemented this simple and broadly applicable protocol as a module within the \textsc{FermiLink} agent harness.\cite{Meng2026} Each optimization begins with a natural-language prompt, according to which the initial source code of the scientific package as well as training and testing inputs are specified (Fig.~\ref{fig:fermilink-optimize-scheme}). A worker agent then iteratively proposes code modifications using the training calculations, while a separate controller accepts or rejects them based on actual performance and scientific accuracy in the testing phase using a different set of testing calculations. As the training and testing computational examples differ from each other, and the worker and controller agents are separated from each other, our approach may reduce outcome bias and lower the risk of overfitting.

In machine learning models, overfitting can occur if datasets are too small.\cite{Hawkins2004,Cawley2010} Since our protocol performs repeated calculations on the same set of chemical systems, more care has to be given to the  preparation of the training and testing calculation examples. In the development of the adaptive SCF method (Sec.~\ref{sec:auto_scf}), for instance, we should improve convergence for challenging systems but retain good performance for more straightforward cases. For achieving this goal, we need to be deliberate in deciding what molecular systems to include in the training and testing. In practice, we included well-behaved calculations as regularization controls alongside the difficult ones for more robust code optimization. 

In addition to overfitting concerns, we must also consider how an agent improves computational efficiency and whether the tradeoffs are acceptable. For example, an agent may explore a wide range of approaches for achieving the optimization goal, some of which may reduce computational time at the expense of substantially increased memory consumption. For this reason, in the Coulomb and exchange matrix optimizations discussed later (Sec.~\ref{sec:lammps_and_pyscf}), we explicitly set memory thresholds to discourage excessive caching.

Furthermore, we must also consider how changing one aspect of the code may influence other functionalities. For instance, in the full-response linear-response time-dependent density functional theory (TDDFT)\cite{Runge1984,Dreuw2005} optimization discussed later (also Sec.~\ref{sec:lammps_and_pyscf}), modifications to the shared preconditioner can affect Casida-TDDFT,\cite{Casida1995} RPA,\cite{Bouman1983} and TDA\cite{Hirata1999} calculations. To make the subsequent optimization more productive, we included Casida calculations alongside full-response calculations to serve as control cases and provided suggested literature on iterative subspace methods.\cite{Olsen1988,Sleijpen1996}

Collectively, the considerations we took during initial preparation for reducing the risk of overfitting and other concerns highlight how scientific insight and human efforts can foster more predictable optimizations. Some limitations may nevertheless emerge only in edge scenarios or when applied to other systems. We therefore performed additional calculations beyond the initial training and testing datasets to assess whether the improvements were more broadly applicable. These findings are presented in the Results and Discussion (Sec.~\ref{sec:results_and_discussion}).

\subsection{Nonbonded Interactions in Classical MD}

Classical MD simulations of biologically relevant systems commonly include explicit water molecules, making water interactions a broadly relevant target for performance optimization. In this work, we consider the TIP4P\cite{Jorgensen1983} family of water models, which, unlike simpler three-site models such as TIP3P\cite{Jorgensen1983} and SPC,\cite{Berendsen1981} places the negative charge on a fourth, virtual site near the oxygen atom. This additional site increases the computational work required to evaluate the electrostatic interactions, making TIP4P simulations generally more computationally demanding than comparable simulations with three-site models. A simple molecular-mechanics energy functional can be expressed as $E_{\mathrm{total}} = E_{\mathrm{bond}} + E_{\mathrm{angle}} + E_{\mathrm{dihedral}} + E_{\mathrm{LJ}} + E_{\mathrm{elec}}$, where the first three terms describe bond stretching, angle bending, and dihedral torsions, respectively. The final two terms represent the nonbonded Lennard--Jones and electrostatic (including both short-range and long-range) interactions.\cite{Senn2009} In practice, the evaluation of these nonbonded interactions often accounts for a substantial fraction of the total simulation time. In LAMMPS,\cite{Thompson2022} the pairwise calculations consider the Lennard--Jones and short-range electrostatic interactions, while the k-space calculations consider the long-range electrostatic contribution. In Sec.~\ref{sec:lammps_and_pyscf}, we focus on improving these two aspects of the LAMMPS code.  

\subsection{SCF Procedure and J/K Interactions}

The common starting point for many popular electronic structure methods is to obtain a set of ground-state molecular orbitals. In Hartree--Fock (HF)\cite{Pople1954,Berthier1954} and Kohn--Sham density functional theory (DFT),\cite{Pople1992} these orbitals are usually expressed as linear combinations of a finite set of Gaussian atomic orbital basis functions, and their expansion coefficients are determined through iterative evaluation of the Roothaan--Hall\cite{Roothaan1951,Hall1951} or Kohn--Sham\cite{Hohenberg1964,Kohn1965} equations, 
\begin{equation}
\mathbf{F}\mathbf{C}
=
\mathbf{S}\mathbf{C}\boldsymbol{\varepsilon},
\label{eq:scf}
\end{equation}
where $\mathbf{F}$, $\mathbf{C}$, $\mathbf{S}$, and $\boldsymbol{\varepsilon}$ are the Fock or Kohn--Sham matrix, orbital coefficient matrix, overlap matrix, and
orbital energy matrix, respectively. Since $\mathbf{F}$ directly depends on the set of orbital coefficients, Eq.~\ref{eq:scf} must be solved iteratively until self-consistency is reached.\cite{Lehtola2020}

During the SCF procedure, construction of $\mathbf{F}$ requires evaluation of the Coulomb matrix $\mathbf{J}$ and, for HF and hybrid DFT, the exact-exchange matrix $
\mathbf{K}$, at each iteration. In restricted HF and DFT, these matrices are obtained by contracting the two-electron, four-center electron-repulsion integrals, $(\mu\nu|\lambda\kappa)$ and $(\mu\lambda|\nu\kappa)$, with an electronic density matrix, $\mathbf{P}$:
\begin{align}
J_{\mu\nu}
&= \sum_{\lambda\kappa}
P_{\lambda\kappa}(\mu\nu|\lambda\kappa)
\label{eq:coulomb}\\
K_{\mu\nu}
&= \sum_{\lambda\kappa}
P_{\lambda\kappa}(\mu\lambda|\nu\kappa).
\label{eq:exchange}
\end{align}
Here, $\mu$, $\nu$, $\lambda$, and $\kappa$ are indices over the electronic basis functions. These contractions become increasingly expensive as the number of basis functions grows, and while there are techniques that can reduce the computational work, repeated J/K matrix construction still accounts for a substantial fraction of the total SCF time. Improving the efficiency of integral evaluations can help reduce the cost of individual iterations in an SCF calculation. We explore this idea in Sec.~\ref{sec:lammps_and_pyscf}.

The overall efficiency of an SCF calculation also depends on the number of iterations required to achieve self-consistency. The direct inversion in the iterative subspace (DIIS) approach\cite{Pulay1980,Pulay1982} is typically the default SCF solver for many electronic structure programs. Its mechanism is to construct the next iteration's Fock matrix through linear combinations of previous $m$ iterations' Fock matrices and choosing expansion coefficients $c_i$ that minimize an SCF error metric:
\begin{equation}
\mathbf{F}_{\mathrm{new}}
=
\sum_{i=1}^{m} c_i \mathbf{F}_i .
\label{eq:diis}
\end{equation}
This solver is often fairly efficient, but it is also highly prone to exhibiting erratic convergence behavior for difficult systems.\cite{Daniels2000,VanVoorhis2002,Chow2026} Second-order methods, such as the co-iterative augmented Hessian (CIAH) method\cite{Sun2016,Sun2017} and the trust-region method\cite{Helmich-Paris2021} implemented in OpenTrustRegion (OTR),\cite{Greiner2026} can sometimes address these convergence difficulties. These methods minimize the electronic energy through direct orbital optimization. The electronic energy can then be approximated to second order as
\begin{equation}
E(\mathbf{x})
\approx
E_0 + \mathbf{g}^{\mathrm{T}}\mathbf{x}
+ \frac{1}{2}\mathbf{x}^{\mathrm{T}}\mathbf{H}\mathbf{x},
\label{eq:sos}
\end{equation}
where $E_0$ is the current iteration's energy, $\mathbf{x}$ are the orbital rotation parameters, and $\mathbf{g}$ and $\mathbf{H}$ are the orbital gradient and Hessian, respectively. These methods require more computational work per SCF iteration and are typically used when DIIS fails. It could therefore be advantageous to have an adaptive SCF strategy that could retain DIIS and automatically detect when a more robust solver might be needed. We explore such a strategy in Sec.~\ref{sec:auto_scf}.

\subsection{Linear-response Time-Dependent DFT}

Many interesting chemical phenomena require a cost-efficient treatment of electronic excited states, which can be accomplished through a response formalism such as TDDFT. In this approach, the Kohn--Sham system is perturbed by a weak, time-dependent external electromagnetic field. This method yields excitation energies and response amplitudes that characterize electronic excitations and other excited state properties. Within the adiabatic approximation, electronic excitation energies $\omega_n$ can be obtained by solving the Casida matrix equation\cite{Casida1995} for $n$ excited states,
\begin{equation}
\begin{pmatrix}
\mathbf{A} & \mathbf{B} \\
\mathbf{B} & \mathbf{A}
\end{pmatrix}
\begin{pmatrix}
\mathbf{X}_n \\
\mathbf{Y}_n
\end{pmatrix}
=
\omega_n
\begin{pmatrix}
\mathbf{I} & \mathbf{0} \\
\mathbf{0} & -\mathbf{I}
\end{pmatrix}
\begin{pmatrix}
\mathbf{X}_n \\
\mathbf{Y}_n
\end{pmatrix}.
\label{eq:tddft}
\end{equation}
Here, the matrices $\mathbf{A}$ and $\mathbf{B}$ contain Coulomb and exchange--correlation response terms, $\mathbf{I}$ is the identity matrix, and $\mathbf{X}_n$ and $\mathbf{Y}_n$ are the excitation and de-excitation amplitudes, respectively. The matrices $\mathbf{A}$ and $\mathbf{B}$ are related to the orbital Hessian that also appears in second-order SCF methods and coupled-perturbed SCF calculations. When only a few excited states are required, a Davidson-type iterative approach\cite{Davidson1975} can be used to solve for the roots of Eq.~\ref{eq:tddft} using repeated Hessian--vector product evaluations, thereby avoiding expensive evaluation of a full analytic orbital Hessian. The Davidson approach can still be expensive and in Sec.~\ref{sec:lammps_and_pyscf}, we further optimize the iterative solver and its corresponding preconditioner to accelerate its convergence.

\section{Computational Details}
\label{sec:computational_details}

After reviewing the computational approaches, we now provide simulation details of the dual-agent optimization approach as well as additional sets of calculations for evaluating the computational code.

\subsection{FermiLink Optimizations}
Code optimizations were performed using the \textsc{FermiLink} harness with OpenAI's\cite{OpenAI2026a} GPT-5.6-sol model, with the exception of the LAMMPS optimizations, which also used the GPT-5.4 and GPT-5.5 models for additional benchmarking. We used the \texttt{xhigh} reasoning effort except during the development of the adaptive SCF strategy where we used the \texttt{max} reasoning effort for both \textsc{FermiLink} and standalone \textsc{Codex}\cite{OpenAI2026b} optimizations. The training and testing datasets along with other optimization settings are described in detail in the Supplementary Material. The LAMMPS optimizations were conducted on the Purdue Anvil HPC cluster using a full compute node comprised of two AMD EPYC 7763 processors. The CPU and GPU optimizations of the serial MD code were conducted on an Apple M4 processor and an NVIDIA RTX 4090 GPU, respectively. All PySCF\cite{Sun2018,Sun2026} optimizations, including the adaptive SCF development, were conducted on an Apple M4 processor. 

\begin{figure}[!t]
\centering
\includegraphics[width=\columnwidth]{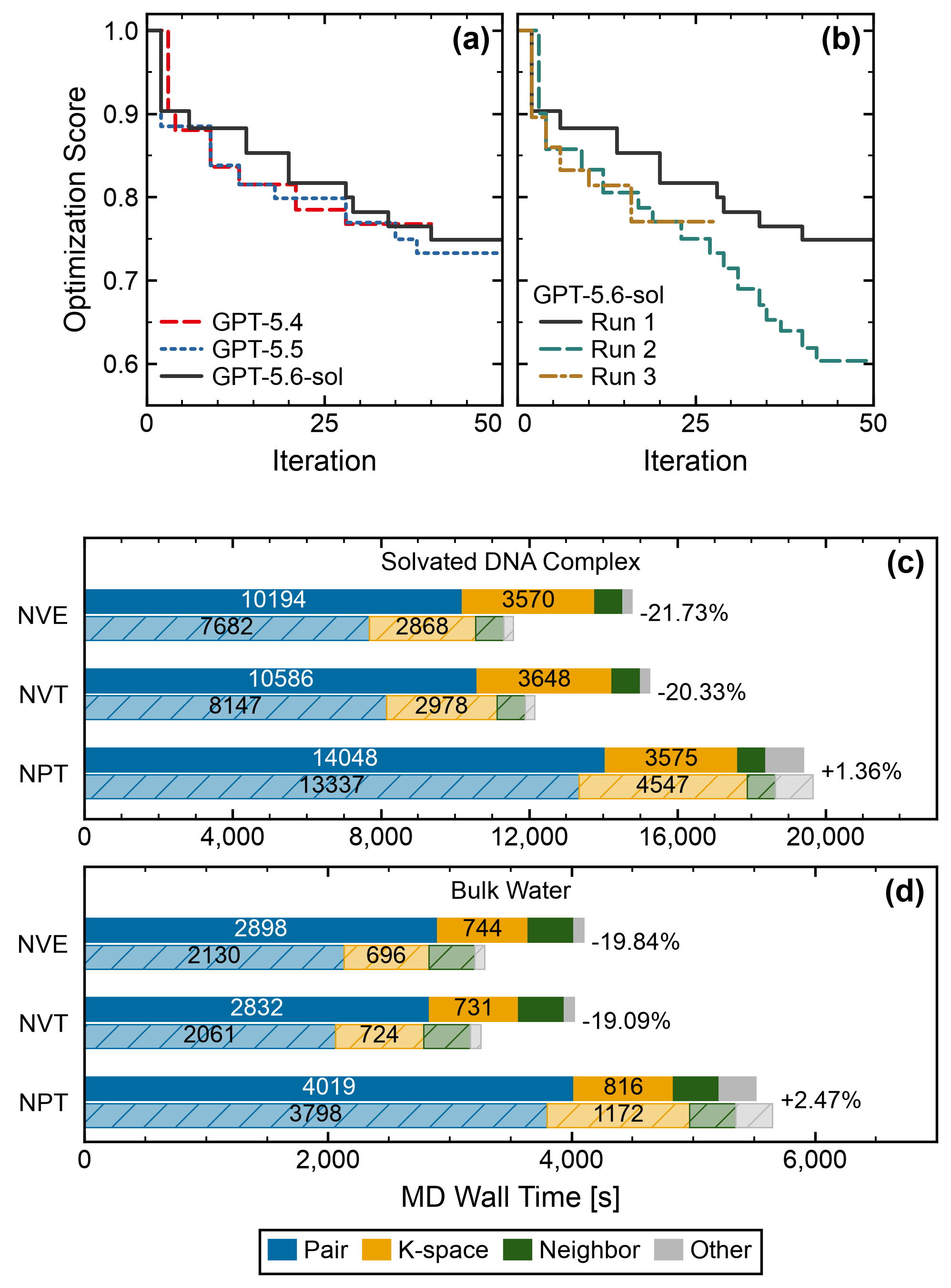}
\caption{
Trajectories for LAMMPS TIP4P water interaction kernel \textsc{FermiLink} optimizations for (a)  multiple runs using different GPT models and (b) repeat runs using the same model. The post-optimization performance comparisons on two systems, a solvated DNA complex (c) and a box of bulk water (d), show that the code optimizations can be applicable to systems beyond training and testing. 
For the \textsc{FermiLink} optimizations in parts (a) and (b), each curve reflects only the accepted code changes. The optimization score reflects the combined pair and k-space evaluation time across test cases relative to the original code, which is assigned a score of one. The GPT-5.6-sol trajectory in panel (a) is the same as Run 1 in panel (b), and the resulting optimized code (same black curve) was used to perform the comparison calculations in panels (c) and (d). In parts (c) and (d), the percentages next to each pair of bar graphs indicate the change in the simulation wall time of the optimized version (lighter, hatched bars) relative to the original version (darker, solid bars).
}
\label{fig:lammps-tip4p-results}
\end{figure}

\begin{figure*}[!t]
\centering
\includegraphics[width=\textwidth]{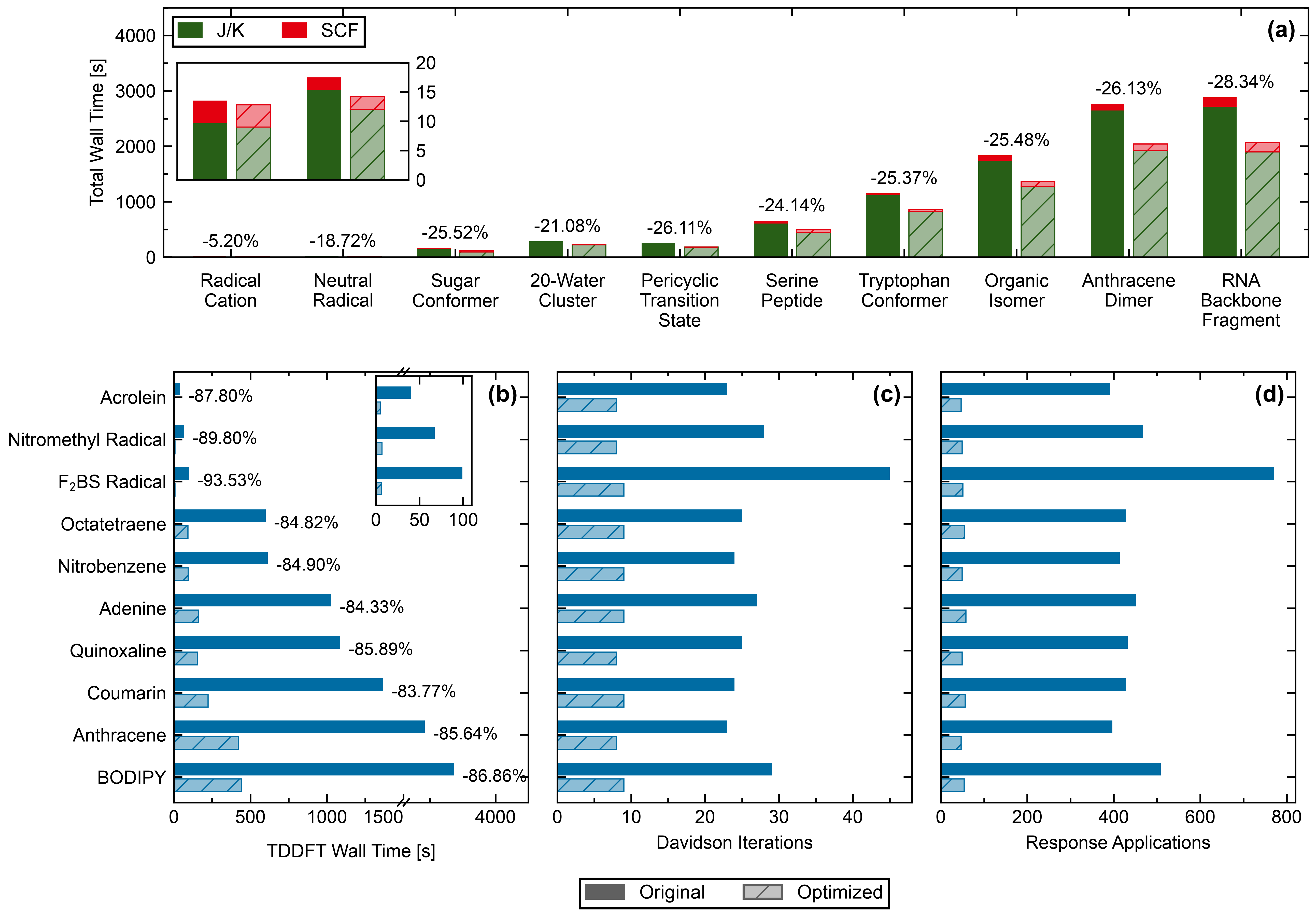}
\caption{
Performance metrics of the original PySCF code compared against the (a)  J/K-optimized and (b-d) TDDFT-optimized  codes. The J/K-optimized code was benchmarked on ten systems selected from the GMTKN55 database.\cite{Goerigk2017} Each pair of bar plots shows the total SCF time (red and green combined) and the total J/K matrix evaluation times as a subset (green). In parts (b-d), the TDDFT-optimized code was benchmarked on ten other systems chosen from the QUESTDB database.\cite{Veril2021,Loos2025} The percentages above or next to each pair of bar graphs indicate the reduction in SCF or TDDFT wall time due to code optimization.
}
\label{fig:pyscf-results}
\end{figure*}

\subsection{Post-Optimization Calculations}

After optimizations, we performed additional calculations using the original and optimized implementations of the codes to assess whether the computational improvements were applicable to additional systems and simulation environments. The original and optimized LAMMPS calculations were conducted on Apple M4 processors using four MPI ranks and one CPU thread per rank. The original and optimized PySCF calculations, which include the J/K and TDDFT calculations along with SCF solver comparisons, were also conducted on Apple M4 processors but used four OpenMP threads. For the CPU and GPU performance comparisons, we used a workstation comprised of an Intel Xeon Silver 4215 processor and a single NVIDIA RTX 4090 GPU. 

For the LAMMPS comparisons in Sec.~\ref{sec:lammps_and_pyscf}, we performed 500 ps NVE, NVT, and NPT simulations on two systems: a solvated DNA complex (PDB: 1BNA),\cite{Drew1981} which was simulated using the Amber OL15 force field,\cite{Zgarbova2015,Galindo-Murillo2016} and a 4,096 molecule box of water. Both systems used the rigid TIP4P-Ew water model\cite{Horn2004} and were
simulated using periodic boundary conditions in all three spatial directions. The NVT and NPT simulations used a Nos\'e--Hoover thermostat\cite{Nose1984,Hoover1985} at 298.15 K, and the NPT simulations used a Nos\'e--Hoover barostat\cite{Hoover1986,Martyna1994} at 1 atm. The DNA complex was simulated using a 1 fs time step, while the bulk water box used 2 fs. The simulations began from identical configurations and velocities following a set of system preparation and equilibration steps as outlined in the Supplementary Material. 

For the CPU- and GPU-optimized study in Sec.~\ref{sec:cpu_and_gpu}, we performed NVE simulations of flexible q-TIP4P/F\cite{Habershon2009} water boxes containing the following number of water molecules: 64, 216, 432, 512, 864, 1,728, 4,096, 8,000, 13,824, 21,952, and 32,768. Each water box was simulated for 1 ps with a 0.5 fs time step and using initial velocities sampled from a Maxwell--Boltzmann distribution at 300 K. Hence, the CPU and GPU calculations used the same initial configuration and velocities for the same sized water box.

In all PySCF assessments in Secs.~\ref{sec:lammps_and_pyscf} and~\ref{sec:auto_scf}, we used the superposition of atomic densities initial guess. The SCF was considered to be converged when the orbital-gradient norm and the energy change between consecutive iterations were below $10^{-4}$ Hartree and $10^{-8}$ Hartree, respectively. For TDDFT calculations, we considered Davidson to be converged when the residual norm is less than $10^{-5}$ Hartree. For the J/K performance evaluation, we assembled a panel of ten molecules obtained from the GMTKN55 database\cite{Goerigk2017} and performed subsequent calculations using both HF and DFT. For TDDFT, we selected ten molecules from the QUESTDB database\cite{Veril2021,Loos2025} and performed calculations using the PBE0 density functional.\cite{Adamo1999} In the SCF solvers panel, we used both HF and DFT for the four systems studied: serine peptide, Fe(NHC-Me)(CO)(NO)$_2$, Fe(CO)$_4$(PH$_3$), and $[\mathrm{NiCl}_4]^{2-}$. All calculations used either the def2-SVP or def2-TZVP electronic basis set.\cite{Weigend2005} The electronic basis set and DFT functional assignment for each system is documented in the Supplementary Material. 

\section{Results and Discussion}
\label{sec:results_and_discussion}

After reviewing the underlying theory and simulation details, we now examine the performance of the \textsc{FermiLink} harness on optimizing and improving computational codes.

\subsection{Optimizations of Established Scientific Packages}
\label{sec:lammps_and_pyscf}

As shown in Fig.~\ref{fig:lammps-tip4p-results}, we first applied optimizations to the TIP4P interaction kernels in LAMMPS using liquid water MD boxes as the training and testing molecular systems. To assess the behavior of our optimization strategy, we performed multiple optimization runs using different GPT models (GPT-5.4, GPT-5.5, and GPT-5.6-sol, Fig.~\ref{fig:lammps-tip4p-results}a) or using the same model multiple times (GPT-5.6-sol, Fig.~\ref{fig:lammps-tip4p-results}b). The optimization score shown in Figs.~\ref{fig:lammps-tip4p-results}a and \ref{fig:lammps-tip4p-results}b is directly related to the combined time for pair and long-range Coulomb evaluations of flexible q-TIP4P/F water simulations. In the GPT model comparison, we found that the optimization score improved by 23--27\%, with GPT-5.5 yielding the lowest final score (Fig.~\ref{fig:lammps-tip4p-results}a). Two additional runs with GPT-5.6-sol showed that the trajectories and final scores can also vary even if the model and optimization conditions are unchanged (Fig.~\ref{fig:lammps-tip4p-results}b). Together, these results highlight the variability in AI agent behavior.

To assess whether these performance enhancements can be extended to other systems and simulation conditions, we applied the GPT-5.6-sol optimized LAMMPS code (black curves) to two additional sets of simulations: a solvated DNA complex and bulk water. These systems used the rigid TIP4P-Ew water model whereas the optimization used flexible q-TIP4P/F water.
The optimized simulations were compared to simulations performed using the original, unmodified LAMMPS code. The original and optimized simulations produced identical trajectories for both systems, but the optimized version reduced the total simulation time by 19--22\% (Figs.~\ref{fig:lammps-tip4p-results}c and \ref{fig:lammps-tip4p-results}d) in most cases. Much of the cost savings arose from 
reductions in the evaluation time of the short-range pair interactions rather than the long-range electrostatic interactions. We found that the performance gains did not apply to pressure-coupled simulations. Interestingly, for these NPT calculations, the k-space evaluation time actually increased as opposed to decreased. Since we performed optimizations in the NVE ensemble, including additional NPT cases into  optimization could yield more favorable performance improvements. Despite this, these results demonstrate that the improvements are extendable to a different water model and to a biomolecular system. 

We next applied the optimization strategy to ground- and excited-state calculations in PySCF. In HF and DFT calculations, iterative evaluation of the electron-repulsion integrals used in the construction of J and K matrices can account for a substantial fraction of the total SCF time. To try and reduce this bottleneck, we optimized components within the \textsc{libcint} integral engine\cite{Sun2015} used in PySCF. For excited-state calculations, we considered TDDFT and focused on improving the convergence of the iterative root search by targeting the Davidson eigensolver and its preconditioner. 

Beginning with the J/K performance assessment, we found that the optimized code reduced the total SCF time by 5--28\% compared to the original code as demonstrated on ten representative systems (Fig.~\ref{fig:pyscf-results}a). We observed that not all calculations devoted the same fraction of time to J/K evaluation. For the two HF examples, the 20-water cluster and the pericyclic transition state, J/K evaluation accounted for nearly all of their SCF time. For the eight DFT calculations, J/K evaluation accounted for slightly less presumably due to the addition of exchange--correlation. In PySCF, J and K integrals can be computed directly on-the-fly, reused by storing them in-core in memory, or use a hybrid in-core/direct approach.\cite{Almlof1982} Since the optimized code specifically targets repetitive J and K integral evaluation, it yielded smaller performance improvements for the radical cation and neutral radical cases, which used an in-core and a hybrid in-core/direct approach, respectively. Larger performance gains were observed for the other eight systems because they used direct integral evaluation.

\begin{figure}[!b]
\centering
\includegraphics[width=\columnwidth]{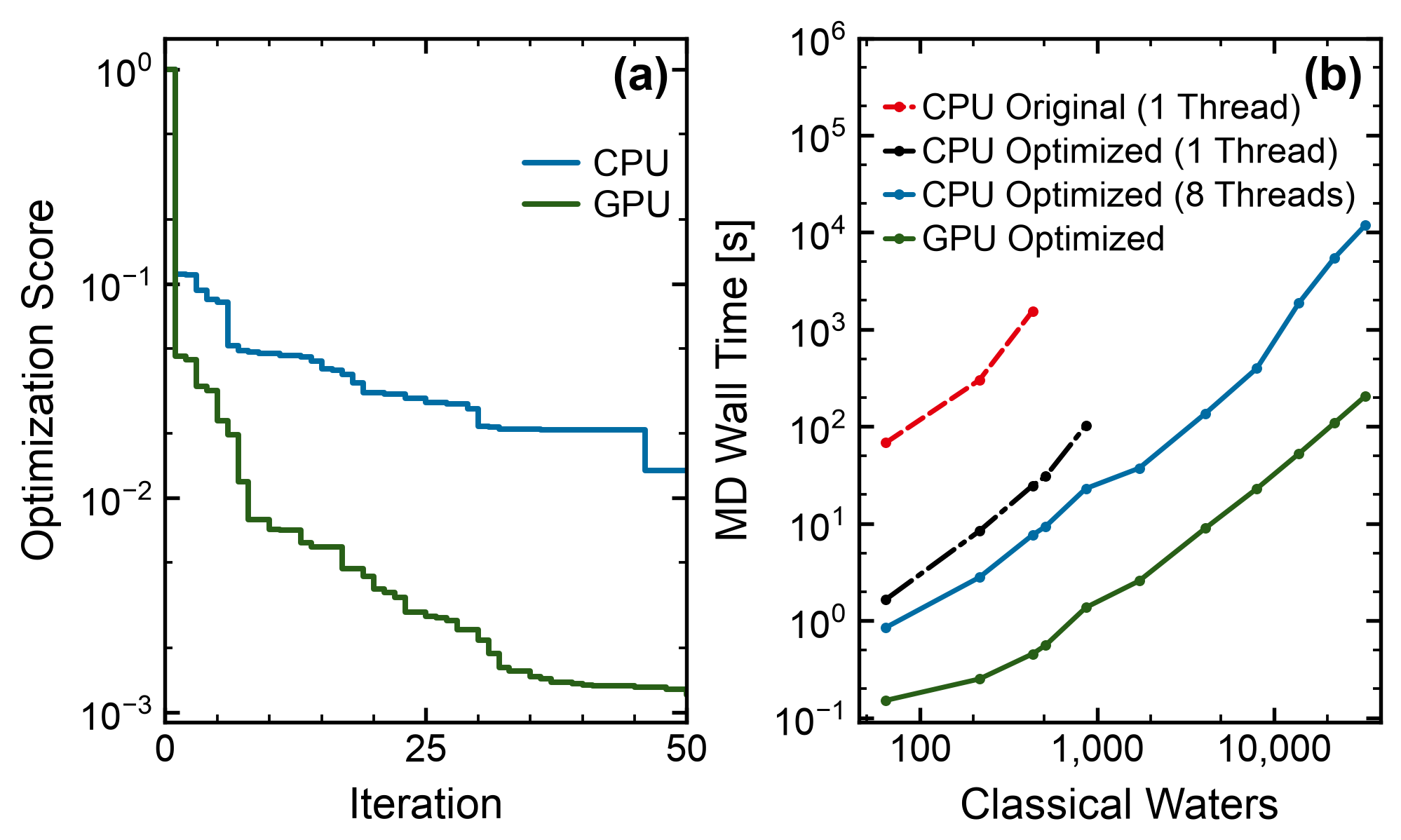}
\caption{
(a) \textsc{FermiLink} optimization trajectories for the independent development of parallel CPU and GPU implementations from the same starting serial MD code. (b) MD simulation times for the original, serial MD code and the optimized codes for a set of flexible q-TIP4P/F water box simulations. Here, the optimization score reflects the relative MD wall times across the test cases, with the original code assigned a score of one. Only a subset of the water box simulations were performed for the single-threaded simulations while the full set was performed for the parallel CPU and GPU versions. 
}
\label{fig:cpu_and_gpu_results}
\end{figure}

Next, we examined the performance of the TDDFT optimized code by applying it to ten representative systems and solving for the first five roots. The optimized calculations yielded nearly identical excitation energies and oscillator strengths as the original calculations (see Supplementary Material) but improved TDDFT time by 84--94\% (Fig.~\ref{fig:pyscf-results}b). This significant cost reduction is due to the optimized code performing far less work per Davidson iteration in addition to requiring fewer iterations overall. 
For instance in F$_2$BS calculations, the number of Davidson iterations decreased from 45 to 9 (Fig.~\ref{fig:pyscf-results}c) and the number of response-matrix applications decreased from 771 to 51 (Fig.~\ref{fig:pyscf-results}d). We identified that the primary improvements in the optimized code compared to the original code are the use of improved preconditioning, a better set of initial guess vectors for the Davidson solver, and a more effective expansion of the Davidson subspace.\cite{Zuev2015,Windom2023,Rappoport2023} 

\begin{figure*}[!t]
\centering
\includegraphics[width=\textwidth]{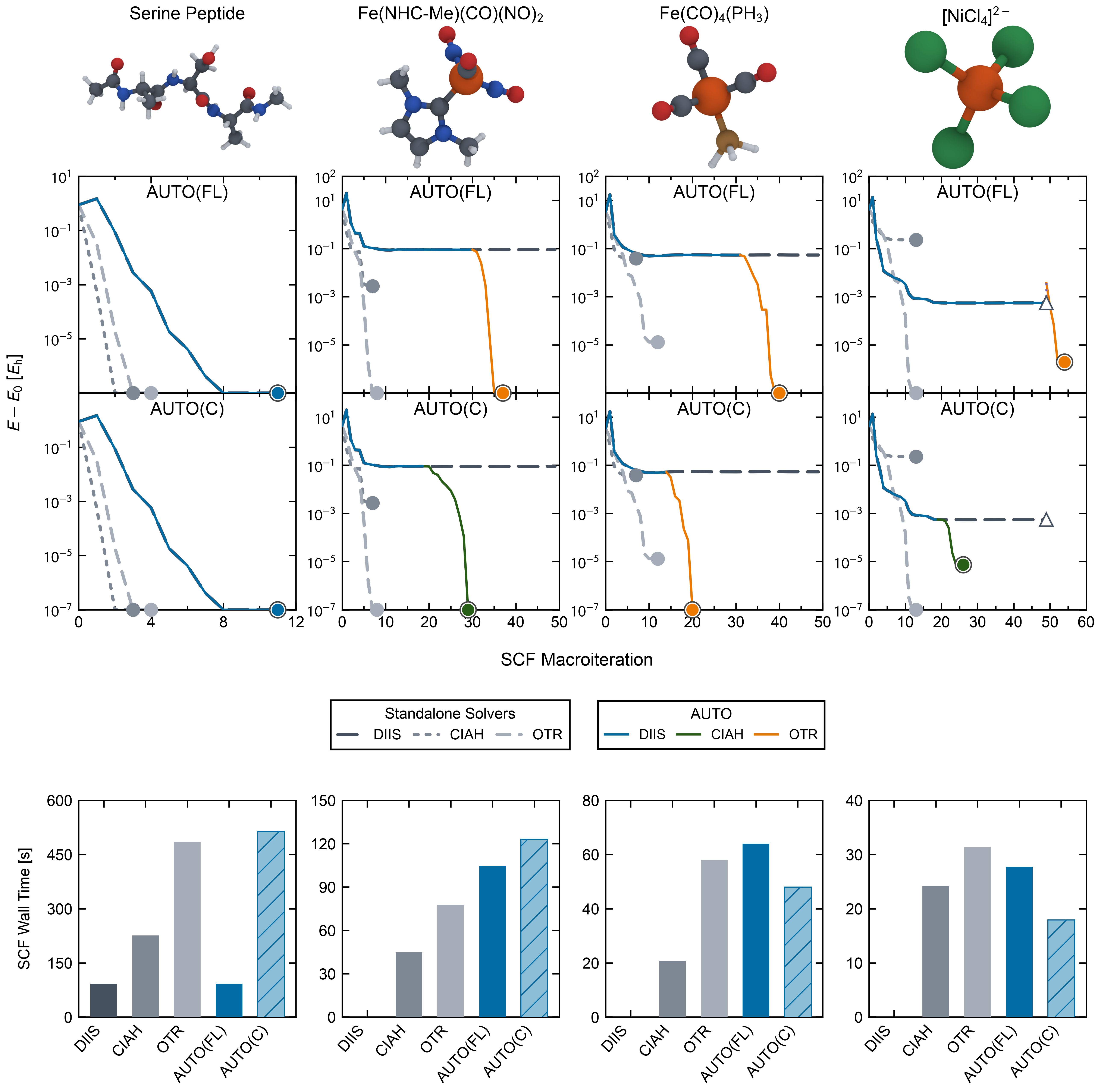}
\caption{%
Four unique molecular systems (top of each column) and their corresponding SCF convergence behavior and convergence times in the adjacent panels below. SCF convergence trajectories for two adaptive SCF implementations, denoted as \texttt{AUTO(FL)} and \texttt{AUTO(C)}, overlay the standalone PySCF solver results depicted in grayscale. $E$ and $E_0$ are the electronic energy at each SCF macroiteration and the lowest-energy solution obtained from the solvers tested, respectively. The filled circle markers represent converged, internally stable solutions while open triangle markers are converged but internally unstable solutions. These endpoint markers were identified following an internal stability analysis of the orbital Hessian after convergence. The reported SCF wall times exclude this separate analysis but include any analysis that was part of the adaptive algorithms. Only the times for converged, internally stable solutions are reported. 
}
\label{fig:adaptive-scf-results}
\end{figure*}

The findings from these J/K and TDDFT calculations are particularly revealing because they indicate two different optimization scopes. In particular, the TDDFT optimizations showcase changes at the algorithmic level that can contribute to reduced iterative work needed to solve the response equations, whereas the J/K optimizations showcase low-level changes including CPU vectorization that improve the overall efficiency of integral evaluations. While there are already established approaches such as density fitting\cite{Whitten1973,Dunlap1979} and Cholesky decomposition\cite{Beebe1977,Koch2003} that can also reduce computational cost for J and K calculations, improving the existing implementation through leaf-level optimizations as demonstrated here offers a complementary choice that requires minimal effort to use and set up and does not change the underlying science.

\subsection{Parallelized CPU and GPU Implementations of a Serial Code}
\label{sec:cpu_and_gpu}

In computational chemistry calculations, beyond using existing scientific packages, very often we need to write our own code for computation. The initial version of the code, while we can work on its scientific accuracy, is usually a serial CPU code. It would be appealing if an agent harness could reliably and significantly improve the speed of this serial CPU code without distorting its accuracy.

To this end, we examined whether we could take a simple, serial MD code and successfully transform it into a scalable and parallelizable implementation for CPUs and GPUs. We performed two independent CPU and GPU optimizations using the same serial CPU code, which we implemented in a development version of the MaxwellLink light-matter simulation package.\cite{Ji2026} For GPU optimization, we prompted the use of mixed single- and double-precision, and for CPU, we emphasized compatibility between both ARM and x86 computer architectures. The optimization trajectories for both are presented in Fig.~\ref{fig:cpu_and_gpu_results}a. Here, the optimization score is directly related to the total MD time of flexible q-TIP4P/F water simulations. These trajectories showcase more steps than the LAMMPS trajectories shown in Fig.~\ref{fig:lammps-tip4p-results}a as they reflect a larger number of accepted code changes. This may be partially because we used a 1\% acceptance threshold (i.e., accepting the change if the speed gain exceeds 1\%) for these optimizations as opposed to the 2\% used for the LAMMPS and PySCF optimizations. We used this lower threshold because we wanted to permit smaller incremental performance gains. We did not rigorously benchmark all combinations and permutations of tunable optimization parameters, but this could be an important future study for investigating optimization pressure and other issues.

After optimization, we applied the optimized CPU and GPU implementations to simulations of the same flexible q-TIP4P/F water model under various system sizes. The results of the optimized codes are compared with the original serial CPU implementation in Fig.~\ref{fig:cpu_and_gpu_results}b. On a single CPU thread, the optimized implementation reduced the total MD time by approximately 1 to 2 orders of magnitude compared to the original implementation. This result indicates that a significant portion of the performance gain is attributed to code re-factorization rather than to parallel execution alone. We performed serial calculations for only several of the smaller water boxes, but these improvements may be even greater for larger water boxes. The GPU implementation provided approximately an order of magnitude improvement over the multithreaded CPU implementation for the smaller systems and almost two orders of magnitude for the larger systems. 

The parallel CPU code was transferable to different computer architectures, but the parallel GPU code only used double-precision despite our suggestion to use mixed-precision. Mixed-precision can further reduce computational cost by evaluating less sensitive contributions (e.g., intermolecular long-range interactions) in single-precision while keeping more sensitive contributions (e.g., intramolecular interactions) in double-precision, as GPUs typically perform single-precision arithmetic faster. These applications show that while AI can generalize the code to different platforms, even with explicit human guidance, it did not incorporate every possible direction that we encouraged. One possible explanation is that mixed precision can introduce numerical deviations over long MD trajectories, which may explain why the agent did not continue in this direction.

\subsection{SCF Algorithm Development by Coordinating Existing Solvers}
\label{sec:auto_scf}

In the preceding sections, we used AI to improve the performance of existing computational methods and implementations. Here, by contrast, we explore its potential for algorithm development by applying the same optimization strategy to develop an approach that could monitor convergence and coordinate different SCF solvers to improve both the robustness and efficiency of SCF calculations. 

We developed such an approach using both \textsc{FermiLink} and standalone \textsc{Codex}, denoted \texttt{AUTO(FL)} and \texttt{AUTO(C)}, respectively. Different from \texttt{AUTO(FL)}, in which the agent iteratively proceeds end-to-end optimization without human intervention, the optimized code obtained using bare \textsc{Codex} relies on iterative streaming interactions between the  human and agent.

As shown in Fig.~\ref{fig:adaptive-scf-results}, we then benchmarked these implementations on four representative systems. These results showcase the SCF convergence behavior for one well-behaved case, serine peptide, and three challenging cases, Fe(NHC-Me)(CO)(NO)$_2$, Fe(CO)$_4$(PH$_3$), and [NiCl$_4$]$^{2-}$. 

For serine peptide, all solvers converged to the same final solution, and both adaptive implementations remained with DIIS throughout the optimization. \texttt{AUTO(FL)} took similar time as standalone DIIS, but \texttt{AUTO(C)} took approximately six times as long because it performed an internal stability analysis after convergence. This automatically triggered analysis was part of the adaptive SCF algorithm and accounted for most of its reported wall time. This additional mechanism might suggest that even with direct human guidance, AI can have the propensity to over-engineer an implementation and substantially increase the computational overhead for an otherwise easy-to-converge calculation.

The three challenging cases allow us to examine how the adaptive implementations respond to convergence difficulties. For both Fe(NHC-Me)(CO)(NO)$_2$ and Fe(CO)$_4$(PH$_3$) systems, standalone DIIS failed to converge, whereas standalone CIAH and OTR converged but to different internally stable solutions. Both adaptive implementations detected a plateau in the orbital-gradient norm and performed different recovery routines. Moreover, both adaptive implementations reached the same lowest-energy, stable solution as OTR for Fe(NHC-Me)(CO)(NO)$_2$. Interestingly, however, both adaptive implementations reached an even lower energy, internally stable solution that was not previously obtained by standalone CIAH or OTR in the case of Fe(CO)$_4$(PH$_3$). For [NiCl$_4$]$^{2-}$, both adaptive implementations converged to internally stable solutions that were slightly higher in energy than the solution obtained with standalone OTR. 

These findings support the presence of multiple stationary points on an SCF energy landscape and showcase how the final solutions can sometimes be sensitive to the SCF optimization path. In fact, the existence of multiple SCF solutions is well documented and has motivated approaches for exploring different solutions.\cite{Gilbert2008,Thom2008,Ye2017,Shea2018} None of the solvers, including both adaptive implementations, can guarantee convergence to a global minimum. It should also be noted that a lower SCF energy does not necessarily imply a more accurate or physical ground state. In the triplet Fe(CO)$_4$(PH$_3$) calculations, the lowest energy solutions (OTR and adaptive) were more spin-contaminated than the higher-energy CIAH solution ($\langle S^2 \rangle \approx 3.326$ versus $2.372$), compared with 2 for a pure triplet. The opposite behavior was observed for triplet [NiCl$_4$]$^{2-}$ in which the lowest energy OTR solution was less spin-contaminated than the higher-energy CIAH solution ($\langle S^2 \rangle \approx 2.002$ versus $3.005$). Our findings directly show that a lower electronic energy can be accompanied either by an increase or decrease in spin-contamination depending on the system. 

Although these adaptive SCF approaches did not consistently reach the lowest observed energy or achieve the shortest wall time, they successfully automated solver selection and performed recovery tactics within a single calculation. This is useful because it reduces the need for users to test different solvers separately. More broadly, these results illustrate how AI can directly translate scientific ideas into working algorithms, with both \textsc{FermiLink} and standalone \textsc{Codex} producing useful adaptive SCF implementations. The two approaches, however, offer different advantages. On the one hand, direct dialogue with standalone \textsc{Codex} allows humans to steer development and refine scientific ideas on-the-fly. On the other hand, \textsc{FermiLink} streamlines the development process using self-sustaining successive rounds of development. Although our present application focused on coordinating existing algorithms, these results motivate further exploration of AI-assisted development that could involve more substantial methodological changes.

\section{Conclusions}
\label{sec:conclusions}

In this work, we explored AI-assisted scientific software and algorithm development using a general training and testing strategy, inspired by the successful practices in machine learning. We extended the approach from accelerating established LAMMPS and PySCF calculations to developing parallel CPU and GPU implementations and an adaptive SCF method. In the TIP4P, J/K, and TDDFT optimizations of established packages, we targeted specific routines of the computational methods. In the CPU and GPU implementations of a serial MD code, the optimization freedom was much greater. In addition, the development of the adaptive SCF approach was the most open-ended and conceptually broadest. This progression showcases a transition from very specific leaf-level changes to more loosely defined targets. Throughout these applications, we relied heavily on our scientific understanding to prepare the optimizations before allowing them to proceed autonomously. As the optimization task became more expansive and less well-defined, it required more human scientific expertise and guidance. Overall, our optimization strategy is one example of a systematic and scalable approach to AI-assisted development. The insights gained from these applications can inform future efforts that combine human scientific expertise with AI to develop new computational methods and explore new chemical perspectives. 

\section*{Supplementary Material}

FermiLink Optimization Details; MD Simulation Details; J/K Calculation Details; TDDFT Calculation Details; Adaptive SCF Calculation Details.

\section*{Acknowledgments}

This material is based upon work supported by the U.S. National Science Foundation under Grant No. CHE-2620630. This work used the Anvil HPC at Purdue University through allocation CHE260067 from the Advanced Cyberinfrastructure Coordination Ecosystem: Services \& Support (ACCESS) program \cite{Boerner2023}, which is supported by U.S. National Science Foundation grants \#2138259, \#2138286, \#2138307, \#2137603, and \#2138296. The authors acknowledge OpenAI for support under the ChatGPT for Academic Researchers program.

\section*{Data Availability}

The data and code used to support the findings in this work will be openly available on Zenodo upon journal publication.

\section*{References}
\bibliography{references}

@article{Meng2026,
  title   = {{FermiLink}: An Agent Harness for Multidomain Scientific Simulations},
  author  = {Meng, Gang and
             Bocanegra Vargas, Andres Felipe and
             Ji, Xinwei and
             Garcia-Gaitan, Federico and
             Reyes-Osorio, Felipe and
             Varela-Manjarres, Jalil and
             Ren, Yafei and
             Dinpajooh, Mohammadhasan and
             Nikoli{\'c}, Branislav K. and
             Li, Tao E.},
  journal = {J. Chem. Theory Comput.},
  year    = {2026},
  note    = {In press}
}

@article{Greiner2026,
title   = {A Reusable Library for Second-Order Orbital Optimization Using the Trust Region Method},
author  = {Greiner, Jonas and
         H{\o}yvik, Ida-Marie and
         Lehtola, Susi and
         Eriksen, Janus J.},
journal = {J. Chem. Theory Comput.},
year    = {2026},
volume  = {22},
number  = {2},
pages   = {881--895},
doi     = {10.1021/acs.jctc.5c01576},
url     = {https://doi.org/10.1021/acs.jctc.5c01576}
}

@article{Helmich-Paris2021,
title   = {A trust-region augmented {Hessian} implementation for restricted and unrestricted {Hartree--Fock} and {Kohn--Sham} methods},
author  = {Helmich-Paris, Benjamin},
journal = {J. Chem. Phys.},
year    = {2021},
volume  = {154},
number  = {16},
pages   = {164104},
doi     = {10.1063/5.0040798},
url     = {https://doi.org/10.1063/5.0040798}
}

@misc{Sun2016,
title         = {Co-iterative augmented {Hessian} method for orbital optimization},
author        = {Sun, Qiming},
year          = {2016},
eprint        = {1610.08423},
archivePrefix = {arXiv},
primaryClass  = {physics.chem-ph},
doi           = {10.48550/arXiv.1610.08423},
url           = {https://arxiv.org/abs/1610.08423}
}

@article{Sun2017,
title   = {A general second order complete active space self-consistent-field solver for large-scale systems},
author  = {Sun, Qiming and
         Yang, Jun and
         Chan, Garnet Kin-Lic},
journal = {Chem. Phys. Lett.},
year    = {2017},
volume  = {683},
pages   = {291--299},
doi     = {10.1016/j.cplett.2017.03.004},
url     = {https://doi.org/10.1016/j.cplett.2017.03.004}
}

@article{Pulay1980,
title   = {Convergence acceleration of iterative sequences. The case of {SCF} iteration},
author  = {Pulay, P{\'e}ter},
journal = {Chem. Phys. Lett.},
year    = {1980},
volume  = {73},
number  = {2},
pages   = {393--398},
doi     = {10.1016/0009-2614(80)80396-4},
url     = {https://doi.org/10.1016/0009-2614(80)80396-4}
}

@article{Pulay1982,
title   = {Improved {SCF} convergence acceleration},
author  = {Pulay, P{\'e}ter},
journal = {J. Comput. Chem.},
year    = {1982},
volume  = {3},
number  = {4},
pages   = {556--560},
doi     = {10.1002/jcc.540030413},
url     = {https://doi.org/10.1002/jcc.540030413}
}

@article{Davidson1975,
title   = {The iterative calculation of a few of the lowest eigenvalues and corresponding eigenvectors of large real-symmetric matrices},
author  = {Davidson, Ernest R.},
journal = {J. Comput. Phys.},
year    = {1975},
volume  = {17},
number  = {1},
pages   = {87--94},
doi     = {10.1016/0021-9991(75)90065-0},
url     = {https://doi.org/10.1016/0021-9991(75)90065-0}
}

@article{Goerigk2017,
title   = {A look at the density functional theory zoo with the advanced {GMTKN55} database for general main group thermochemistry, kinetics and noncovalent interactions},
author  = {Goerigk, Lars and
         Hansen, Andreas and
         Bauer, Christoph and
         Ehrlich, Stephan and
         Najibi, Asim and
         Grimme, Stefan},
journal = {Phys. Chem. Chem. Phys.},
year    = {2017},
volume  = {19},
number  = {48},
pages   = {32184--32215},
doi     = {10.1039/c7cp04913g},
url     = {https://doi.org/10.1039/c7cp04913g}
}

@article{Loos2025,
title   = {{QUEST} Database of Highly-Accurate Excitation Energies},
author  = {Loos, Pierre-Fran{\c{c}}ois and
         Boggio-Pasqua, Martial and
         Blondel, Aymeric and
         Lipparini, Filippo and
         Jacquemin, Denis},
journal = {J. Chem. Theory Comput.},
year    = {2025},
volume  = {21},
number  = {16},
pages   = {8010--8033},
doi     = {10.1021/acs.jctc.5c00975},
url     = {https://doi.org/10.1021/acs.jctc.5c00975}
}

@article{Veril2021,
title   = {{QUESTDB}: A database of highly accurate excitation energies for the electronic structure community},
author  = {V{\'e}ril, Micka{\"e}l and
         Scemama, Anthony and
         Caffarel, Michel and
         Lipparini, Filippo and
         Boggio-Pasqua, Martial and
         Jacquemin, Denis and
         Loos, Pierre-Fran{\c{c}}ois},
journal = {WIREs Comput. Mol. Sci.},
year    = {2021},
volume  = {11},
number  = {5},
pages   = {e1517},
doi     = {10.1002/wcms.1517},
url     = {https://doi.org/10.1002/wcms.1517}
}

@article{Thompson2022,
title   = {{LAMMPS} - a flexible simulation tool for particle-based materials modeling at the atomic, meso, and continuum scales},
author  = {Thompson, Aidan P. and
         Aktulga, H. Metin and
         Berger, Richard and
         Bolintineanu, Dan S. and
         Brown, W. Michael and
         Crozier, Paul S. and
         in 't Veld, Pieter J. and
         Kohlmeyer, Axel and
         Moore, Stan G. and
         Nguyen, Trung Dac and
         Shan, Ray and
         Stevens, Mark J. and
         Tranchida, Julien and
         Trott, Christian and
         Plimpton, Steven J.},
journal = {Comput. Phys. Commun.},
year    = {2022},
volume  = {271},
pages   = {108171},
doi     = {10.1016/j.cpc.2021.108171},
url     = {https://doi.org/10.1016/j.cpc.2021.108171}
}

@article{Sun2026,
title   = {The {Python} simulations of chemistry framework: 10 years of an open-source quantum chemistry project},
author  = {Sun, Qiming and
         Hermes, Matthew R. and
         Wu, Xiaojie and
         Zhai, Huanchen and
         Zhang, Xing and
         Ahmed, Abdelrahman M. and
         Aucar, Juan Jos{\'e} and
         Backhouse, Oliver J. and
         Banerjee, Samragni and
         Bao, Peng and
         Bogdanov, Nikolay A. and
         Bystrom, Kyle and
         Chapoton, Fr{\'e}d{\'e}ric and
         Chen, Ning-Yuan and
         Chernyshov, Ivan Yu. and
         Clifford, Helen S. and
         Cohen-Janes, Sander and
         Cui, Zhi-Hao and
         Damour, Yann D. and
         Dattani, Nike and
         Dittmer, Linus Bjarne and
         Ehlert, Sebastian and
         Eriksen, Janus Juul and
         Evangelista, Francesco A. and
         Ewing, Simon A. and
         Farahvash, Ardavan and
         Focke, Kevin and
         Gao, Yang and
         Gasperich, Kevin E. and
         Gillispie, Nathan and
         Greiner, Jonas and
         Hennefarth, Matthew R. and
         Hermann, Jan and
         Hillenbrand, Christopher and
         Huhtasalo, Joonatan and
         Ibrahim, Basil and
         Jangid, Bhavnesh and
         Javaremi, Alireza Nejati and
         Jenkins, Andrew J. and
         Jin, Yu and
         King, Daniel S. and
         Kooi, Derk Pieter and
         Kurian, Jo S. and
         Larsson, Henrik R. and
         Lau, Bryan Tak Gwong and
         Lee, Seunghoon and
         Lehtola, Susi and
         Li, Chenghan and
         Li, Hao and
         Li, Jiachen and
         Li, Rui and
         Li, Shuhang and
         Lykhin, Aleksandr O. and
         Mahajan, Ankit and
         Mauger, Nastasia and
         del Mazo-Sevillano, Pablo and
         Moussa, Jonathan and
         Nakano, Kousuke and
         Neufeld, Verena A. and
         Peng, Linqing and
         Pham, Hung Q. and
         Pinski, Peter and
         Pokhilko, Pavel and
         Pu, Zhichen and
         Qian, Yubing and
         Quiton, Stephen Jon and
         Schulze, Wanja T. and
         Scott, Thais R. and
         Seal, Aniruddha and
         Serna, James D. and
         Smith, James E. T. and
         Smyser, Kori E. and
         Stahl, Terrence and
         Sun, Chong and
         Sung, Kevin J. and
         Trushin, Egor and
         Upadhyay, Shiv and
         Vo, Ethan A. and
         Vogels, Thijs and
         Wang, Shirong and
         Wang, Tai and
         Wang, Xiao and
         Wang, Xubo and
         Wang, Yuanheng and
         Williamson, Mark and
         Yang, Junjie and
         Ye, Hong-Zhou and
         Yeh, Chia-Nan and
         Yu, Haiyang and
         Yu, Jincheng and
         Yu, Victor Wen-zhe and
         Zhang, Chaoqun and
         Zhang, Dayou and
         Zhang, Yichi and
         Zhao, Zijun and
         Zhou, Zehao and
         Zhu, Andrew J. and
         Zhu, Tianyu and
         Berkelbach, Timothy C. and
         Gagliardi, Laura and
         Sharma, Sandeep and
         Sokolov, Alexander Yu. and
         Chan, Garnet Kin-Lic},
journal = {J. Chem. Phys.},
year    = {2026},
volume  = {165},
number  = {10},
pages   = {102502},
doi     = {10.1063/5.0337441},
url     = {https://doi.org/10.1063/5.0337441}
}

@article{Sun2018,
title   = {{PySCF}: the {Python}-based simulations of chemistry framework},
author  = {Sun, Qiming and
         Berkelbach, Timothy C. and
         Blunt, Nick S. and
         Booth, George H. and
         Guo, Sheng and
         Li, Zhendong and
         Liu, Junzi and
         McClain, James D. and
         Sayfutyarova, Elvira R. and
         Sharma, Sandeep and
         Wouters, Sebastian and
         Chan, Garnet Kin-Lic},
journal = {WIREs Comput. Mol. Sci.},
year    = {2018},
volume  = {8},
number  = {1},
pages   = {e1340},
doi     = {10.1002/wcms.1340},
url     = {https://doi.org/10.1002/wcms.1340}
}

@article{Sun2015,
title   = {{Libcint}: An efficient general integral library for {Gaussian} basis functions},
author  = {Sun, Qiming},
journal = {J. Comput. Chem.},
year    = {2015},
volume  = {36},
number  = {22},
pages   = {1664--1671},
doi     = {10.1002/jcc.23981},
url     = {https://doi.org/10.1002/jcc.23981}
}

@article{Jorgensen1983,
title   = {Comparison of simple potential functions for simulating liquid water},
author  = {Jorgensen, William L. and Chandrasekhar, Jayaraman and Madura, Jeffry D. and Impey, Roger W. and Klein, Michael L.},
journal = {J. Chem. Phys.},
year    = {1983},
volume  = {79},
number  = {2},
pages   = {926--935},
doi     = {10.1063/1.445869},
url     = {https://doi.org/10.1063/1.445869}
}

@article{Habershon2009,
title   = {Competing quantum effects in the dynamics of a flexible water model},
author  = {Habershon, Scott and Markland, Thomas E. and Manolopoulos, David E.},
journal = {J. Chem. Phys.},
year    = {2009},
volume  = {131},
number  = {2},
pages   = {024501},
doi     = {10.1063/1.3167790},
url     = {https://doi.org/10.1063/1.3167790}
}

@article{Horn2004,
title   = {Development of an improved four-site water model for biomolecular simulations: {TIP4P-Ew}},
author  = {Horn, Hans W. and Swope, William C. and Pitera, Jed W. and Madura, Jeffry D. and Dick, Thomas J. and Hura, Greg L. and Head-Gordon, Teresa},
journal = {J. Chem. Phys.},
year    = {2004},
volume  = {120},
number  = {20},
pages   = {9665--9678},
doi     = {10.1063/1.1683075},
url     = {https://doi.org/10.1063/1.1683075}
}

@incollection{Berendsen1981,
title     = {Interaction Models for Water in Relation to Protein Hydration},
author    = {Berendsen, H. J. C. and Postma, J. P. M. and van Gunsteren, W. F. and Hermans, J.},
editor    = {Pullman, Bernard},
booktitle = {Intermolecular Forces},
series    = {The Jerusalem Symposia on Quantum Chemistry and Biochemistry},
volume    = {14},
year      = {1981},
pages     = {331--342},
publisher = {Springer},
address   = {Dordrecht},
doi       = {10.1007/978-94-015-7658-1_21},
url       = {https://doi.org/10.1007/978-94-015-7658-1_21}
}

@article{Hoover1985,
title   = {Canonical dynamics: Equilibrium phase-space distributions},
author  = {Hoover, William G.},
journal = {Phys. Rev. A},
year    = {1985},
volume  = {31},
number  = {3},
pages   = {1695--1697},
doi     = {10.1103/PhysRevA.31.1695},
url     = {https://doi.org/10.1103/PhysRevA.31.1695}
}

@article{Martyna1994,
title   = {Constant pressure molecular dynamics algorithms},
author  = {Martyna, Glenn J. and Tobias, Douglas J. and Klein, Michael L.},
journal = {J. Chem. Phys.},
year    = {1994},
volume  = {101},
number  = {5},
pages   = {4177--4189},
doi     = {10.1063/1.467468},
url     = {https://doi.org/10.1063/1.467468}
}

@article{Nose1984,
title   = {A molecular dynamics method for simulations in the canonical ensemble},
author  = {Nos{\'e}, Shuichi},
journal = {Mol. Phys.},
year    = {1984},
volume  = {52},
number  = {2},
pages   = {255--268},
doi     = {10.1080/00268978400101201},
url     = {https://doi.org/10.1080/00268978400101201}
}

@article{Hoover1986,
title   = {Constant-pressure equations of motion},
author  = {Hoover, William G.},
journal = {Phys. Rev. A},
year    = {1986},
volume  = {34},
number  = {3},
pages   = {2499--2500},
doi     = {10.1103/PhysRevA.34.2499},
url     = {https://doi.org/10.1103/PhysRevA.34.2499}
}

@article{Ji2026,
title   = {{MaxwellLink}: A Unified Framework for Self-Consistent Light--Matter Simulations},
author  = {Ji, Xinwei and Bocanegra Vargas, Andres Felipe and Meng, Gang and Li, Tao E.},
journal = {J. Chem. Theory Comput.},
year    = {2026},
volume  = {22},
number  = {6},
pages   = {2725--2738},
doi     = {10.1021/acs.jctc.5c02028},
url     = {https://doi.org/10.1021/acs.jctc.5c02028}
}

@misc{OpenAI2026a,
title        = {All Models},
author       = {{OpenAI}},
year         = {2026},
howpublished = {OpenAI API documentation},
url          = {https://developers.openai.com/api/docs/models/all},
note         = {Accessed October 6, 2026}
}

@misc{OpenAI2026b,
title        = {{Codex CLI}},
author       = {{OpenAI}},
year         = {2026},
howpublished = {OpenAI documentation},
url          = {https://learn.chatgpt.com/docs/codex/cli},
note         = {Accessed October 6, 2026}
}

@article{Runge1984,
title   = {Density-Functional Theory for Time-Dependent Systems},
author  = {Runge, Erich and Gross, E. K. U.},
journal = {Phys. Rev. Lett.},
year    = {1984},
volume  = {52},
number  = {12},
pages   = {997--1000},
doi     = {10.1103/PhysRevLett.52.997},
url     = {https://doi.org/10.1103/PhysRevLett.52.997}
}

@incollection{Casida1995,
title     = {Time-Dependent Density Functional Response Theory for Molecules},
author    = {Casida, Mark E.},
editor    = {Chong, D. P.},
booktitle = {Recent Advances in Density Functional Methods, Part I},
year      = {1995},
pages     = {155--192},
publisher = {World Scientific},
address   = {Singapore},
doi       = {10.1142/9789812830586_0005},
url       = {https://doi.org/10.1142/9789812830586_0005}
}

@article{Dreuw2005,
title   = {Single-Reference ab Initio Methods for the Calculation of Excited States of Large Molecules},
author  = {Dreuw, Andreas and Head-Gordon, Martin},
journal = {Chem. Rev.},
year    = {2005},
volume  = {105},
number  = {11},
pages   = {4009--4037},
doi     = {10.1021/cr0505627},
url     = {https://doi.org/10.1021/cr0505627}
}

@article{Hirata1999,
title   = {Time-dependent density functional theory within the {Tamm--Dancoff} approximation},
author  = {Hirata, So and Head-Gordon, Martin},
journal = {Chem. Phys. Lett.},
year    = {1999},
volume  = {314},
number  = {3--4},
pages   = {291--299},
doi     = {10.1016/S0009-2614(99)01149-5},
url     = {https://doi.org/10.1016/S0009-2614(99)01149-5}
}

@article{Bouman1983,
title   = {Large-scale {RPA} calculations of chiroptical properties of organic molecules: Program {RPAC}},
author  = {Bouman, Thomas D. and Hansen, Aage E. and Voigt, Bj{\o}rn and Rettrup, Sten},
journal = {Int. J. Quantum Chem.},
year    = {1983},
volume  = {23},
number  = {2},
pages   = {595--611},
doi     = {10.1002/qua.560230230},
url     = {https://doi.org/10.1002/qua.560230230}
}

@article{Senn2009,
title   = {{QM/MM} Methods for Biomolecular Systems},
author  = {Senn, Hans Martin and Thiel, Walter},
journal = {Angew. Chem. Int. Ed.},
year    = {2009},
volume  = {48},
number  = {7},
pages   = {1198--1229},
doi     = {10.1002/anie.200802019},
url     = {https://doi.org/10.1002/anie.200802019}
}

@article{Zgarbova2015,
title   = {Refinement of the Sugar--Phosphate Backbone Torsion Beta for {AMBER} Force Fields Improves the Description of {Z- and B-DNA}},
author  = {Zgarbov{\'a}, Marie and
         {\v{S}}poner, Ji{\v{r}}{\'i} and
         Otyepka, Michal and
         Cheatham, III, Thomas E. and
         Galindo-Murillo, Rodrigo and
         Jure{\v{c}}ka, Petr},
journal = {J. Chem. Theory Comput.},
year    = {2015},
volume  = {11},
number  = {12},
pages   = {5723--5736},
doi     = {10.1021/acs.jctc.5b00716},
url     = {https://doi.org/10.1021/acs.jctc.5b00716}
}

@article{Galindo-Murillo2016,
title   = {Assessing the Current State of {Amber} Force Field Modifications for {DNA}},
author  = {Galindo-Murillo, Rodrigo and
         Robertson, James C. and
         Zgarbov{\'a}, Marie and
         {\v{S}}poner, Ji{\v{r}}{\'i} and
         Otyepka, Michal and
         Jure{\v{c}}ka, Petr and
         Cheatham, III, Thomas E.},
journal = {J. Chem. Theory Comput.},
year    = {2016},
volume  = {12},
number  = {8},
pages   = {4114--4127},
doi     = {10.1021/acs.jctc.6b00186},
url     = {https://doi.org/10.1021/acs.jctc.6b00186}
}

@article{Drew1981,
title   = {Structure of a {B-DNA} dodecamer: conformation and dynamics},
author  = {Drew, H. R. and
         Wing, R. M. and
         Takano, T. and
         Broka, C. and
         Tanaka, S. and
         Itakura, K. and
         Dickerson, R. E.},
journal = {Proc. Natl. Acad. Sci. U.S.A.},
year    = {1981},
volume  = {78},
number  = {4},
pages   = {2179--2183},
doi     = {10.1073/pnas.78.4.2179},
url     = {https://doi.org/10.1073/pnas.78.4.2179}
}

@article{Gilbert2008,
title   = {Self-Consistent Field Calculations of Excited States Using the Maximum Overlap Method ({MOM})},
author  = {Gilbert, Andrew T. B. and Besley, Nicholas A. and Gill, Peter M. W.},
journal = {J. Phys. Chem. A},
year    = {2008},
volume  = {112},
number  = {50},
pages   = {13164--13171},
doi     = {10.1021/jp801738f},
url     = {https://doi.org/10.1021/jp801738f}
}

@article{Thom2008,
title   = {Locating Multiple Self-Consistent Field Solutions: An Approach Inspired by Metadynamics},
author  = {Thom, Alex J. W. and Head-Gordon, Martin},
journal = {Phys. Rev. Lett.},
year    = {2008},
volume  = {101},
number  = {19},
pages   = {193001},
doi     = {10.1103/PhysRevLett.101.193001},
url     = {https://doi.org/10.1103/PhysRevLett.101.193001}
}

@article{Ye2017,
title   = {{$\sigma$-SCF}: A direct energy-targeting method to mean-field excited states},
author  = {Ye, Hong-Zhou and Welborn, Matthew and Ricke, Nathan D. and Van Voorhis, Troy},
journal = {J. Chem. Phys.},
year    = {2017},
volume  = {147},
number  = {21},
pages   = {214104},
doi     = {10.1063/1.5001262},
url     = {https://doi.org/10.1063/1.5001262}
}

@article{Shea2018,
title   = {Communication: A mean field platform for excited state quantum chemistry},
author  = {Shea, Jacqueline A. R. and Neuscamman, Eric},
journal = {J. Chem. Phys.},
year    = {2018},
volume  = {149},
number  = {8},
pages   = {081101},
doi     = {10.1063/1.5045056},
url     = {https://doi.org/10.1063/1.5045056}
}

@article{Chow2026,
title   = {Efficient and robust optimization of nuclear and electronic orbitals within the nuclear {Hartree} product representation},
author  = {Chow, Mathew and Paenurk, Eno and Hammes-Schiffer, Sharon},
journal = {J. Chem. Phys.},
year    = {2026},
volume  = {164},
number  = {7},
pages   = {074103},
doi     = {10.1063/5.0308634},
url     = {https://doi.org/10.1063/5.0308634}
}

@article{Daniels2000,
title   = {Converging difficult {SCF} cases with conjugate gradient density matrix search},
author  = {Daniels, Andrew D. and Scuseria, Gustavo E.},
journal = {Phys. Chem. Chem. Phys.},
year    = {2000},
volume  = {2},
number  = {10},
pages   = {2173--2176},
doi     = {10.1039/B000618L},
url     = {https://doi.org/10.1039/B000618L}
}

@article{VanVoorhis2002,
title   = {A geometric approach to direct minimization},
author  = {Van Voorhis, Troy and Head-Gordon, Martin},
journal = {Mol. Phys.},
year    = {2002},
volume  = {100},
number  = {11},
pages   = {1713--1721},
doi     = {10.1080/00268970110103642},
url     = {https://doi.org/10.1080/00268970110103642}
}

@article{Whitten1973,
title   = {Coulombic potential energy integrals and approximations},
author  = {Whitten, J. L.},
journal = {J. Chem. Phys.},
year    = {1973},
volume  = {58},
number  = {10},
pages   = {4496--4501},
doi     = {10.1063/1.1679012},
url     = {https://doi.org/10.1063/1.1679012}
}

@article{Dunlap1979,
title   = {On some approximations in applications of {$X\alpha$} theory},
author  = {Dunlap, B. I. and Connolly, J. W. D. and Sabin, J. R.},
journal = {J. Chem. Phys.},
year    = {1979},
volume  = {71},
number  = {8},
pages   = {3396--3402},
doi     = {10.1063/1.438728},
url     = {https://doi.org/10.1063/1.438728}
}

@article{Beebe1977,
title   = {Simplifications in the generation and transformation of two-electron integrals in molecular calculations},
author  = {Beebe, Nelson H. F. and Linderberg, Jan},
journal = {Int. J. Quantum Chem.},
year    = {1977},
volume  = {12},
number  = {4},
pages   = {683--705},
doi     = {10.1002/qua.560120408},
url     = {https://doi.org/10.1002/qua.560120408}
}

@article{Koch2003,
title   = {Reduced scaling in electronic structure calculations using {Cholesky} decompositions},
author  = {Koch, Henrik and S{\'a}nchez de Mer{\'a}s, Alfredo and Pedersen, Thomas Bondo},
journal = {J. Chem. Phys.},
year    = {2003},
volume  = {118},
number  = {21},
pages   = {9481--9484},
doi     = {10.1063/1.1578621},
url     = {https://doi.org/10.1063/1.1578621}
}

@article{Rappoport2023,
title   = {{Libkrylov}: A modular open-source software library for extremely large on-the-fly matrix computations},
author  = {Rappoport, Dmitrij and Bekoe, Samuel and Mohanam, Luke Nambi and
         Le, Scott and George, Naje' and Shen, Ziyue and Furche, Filipp},
journal = {J. Comput. Chem.},
year    = {2023},
volume  = {44},
number  = {11},
pages   = {1105--1118},
doi     = {10.1002/jcc.27068},
url     = {https://doi.org/10.1002/jcc.27068}
}

@article{Olsen1988,
title   = {Solution of the large matrix equations which occur in response theory},
author  = {Olsen, Jeppe and Jensen, Hans J{\o}rgen Aa. and J{\o}rgensen, Poul},
journal = {J. Comput. Phys.},
year    = {1988},
volume  = {74},
number  = {2},
pages   = {265--282},
doi     = {10.1016/0021-9991(88)90081-2},
url     = {https://doi.org/10.1016/0021-9991(88)90081-2}
}

@article{Zuev2015,
title   = {New algorithms for iterative matrix-free eigensolvers in quantum chemistry},
author  = {Zuev, Dmitry and Vecharynski, Eugene and Yang, Chao and
         Orms, Natalie and Krylov, Anna I.},
journal = {J. Comput. Chem.},
year    = {2015},
volume  = {36},
number  = {5},
pages   = {273--284},
doi     = {10.1002/jcc.23800},
url     = {https://doi.org/10.1002/jcc.23800}
}

@article{Windom2023,
title   = {On the iterative diagonalization of matrices in quantum chemistry: Reconciling preconditioner design with {Brillouin--Wigner} perturbation theory},
author  = {Windom, Zachary W. and Bartlett, Rodney J.},
journal = {J. Chem. Phys.},
year    = {2023},
volume  = {158},
number  = {13},
pages   = {134107},
doi     = {10.1063/5.0139295},
url     = {https://doi.org/10.1063/5.0139295}
}

@article{Sleijpen1996,
title   = {A {Jacobi--Davidson} Iteration Method for Linear Eigenvalue Problems},
author  = {Sleijpen, Gerard L. G. and van der Vorst, Henk A.},
journal = {SIAM J. Matrix Anal. Appl.},
year    = {1996},
volume  = {17},
number  = {2},
pages   = {401--425},
doi     = {10.1137/S0895479894270427},
url     = {https://doi.org/10.1137/S0895479894270427}
}

@article{Roothaan1951,
title   = {New Developments in Molecular Orbital Theory},
author  = {Roothaan, C. C. J.},
journal = {Rev. Mod. Phys.},
year    = {1951},
volume  = {23},
number  = {2},
pages   = {69--89},
doi     = {10.1103/RevModPhys.23.69},
url     = {https://doi.org/10.1103/RevModPhys.23.69}
}

@article{Hall1951,
title   = {The molecular orbital theory of chemical valency {VIII}. A method of calculating ionization potentials},
author  = {Hall, G. G.},
journal = {Proc. R. Soc. Lond. A},
year    = {1951},
volume  = {205},
number  = {1083},
pages   = {541--552},
doi     = {10.1098/rspa.1951.0048},
url     = {https://doi.org/10.1098/rspa.1951.0048}
}

@article{Hohenberg1964,
title   = {Inhomogeneous Electron Gas},
author  = {Hohenberg, P. and Kohn, W.},
journal = {Phys. Rev.},
year    = {1964},
volume  = {136},
number  = {3B},
pages   = {B864--B871},
doi     = {10.1103/PhysRev.136.B864},
url     = {https://doi.org/10.1103/PhysRev.136.B864}
}

@article{Kohn1965,
title   = {Self-Consistent Equations Including Exchange and Correlation Effects},
author  = {Kohn, W. and Sham, L. J.},
journal = {Phys. Rev.},
year    = {1965},
volume  = {140},
number  = {4A},
pages   = {A1133--A1138},
doi     = {10.1103/PhysRev.140.A1133},
url     = {https://doi.org/10.1103/PhysRev.140.A1133}
}

@article{Pople1954,
title   = {Self-Consistent Orbitals for Radicals},
author  = {Pople, J. A. and Nesbet, R. K.},
journal = {J. Chem. Phys.},
year    = {1954},
volume  = {22},
number  = {3},
pages   = {571--572},
doi     = {10.1063/1.1740120},
url     = {https://doi.org/10.1063/1.1740120}
}

@article{Berthier1954,
title   = {Configurations {\'e}lectroniques incompl{\`e}tes},
author  = {Berthier, Gaston},
journal = {J. Chim. Phys.},
year    = {1954},
volume  = {51},
pages   = {363--371},
doi     = {10.1051/jcp/1954510363},
url     = {https://doi.org/10.1051/jcp/1954510363}
}

@article{Pople1992,
title   = {{Kohn--Sham} density-functional theory within a finite basis set},
author  = {Pople, John A. and Gill, Peter M. W. and Johnson, Benny G.},
journal = {Chem. Phys. Lett.},
year    = {1992},
volume  = {199},
number  = {6},
pages   = {557--560},
doi     = {10.1016/0009-2614(92)85009-Y},
url     = {https://doi.org/10.1016/0009-2614(92)85009-Y}
}

@article{Almlof1982,
title   = {Principles for a direct {SCF} approach to {LICAO--MO} ab-initio calculations},
author  = {Alml{\"o}f, J. and Faegri, Jr., K. and Korsell, K.},
journal = {J. Comput. Chem.},
year    = {1982},
volume  = {3},
number  = {3},
pages   = {385--399},
doi     = {10.1002/jcc.540030314},
url     = {https://doi.org/10.1002/jcc.540030314}
}

@article{Lehtola2020,
title   = {An Overview of Self-Consistent Field Calculations Within Finite Basis Sets},
author  = {Lehtola, Susi and Blockhuys, Frank and Van Alsenoy, Christian},
journal = {Molecules},
year    = {2020},
volume  = {25},
number  = {5},
pages   = {1218},
doi     = {10.3390/molecules25051218},
url     = {https://doi.org/10.3390/molecules25051218}
}

@article{LeCun2015,
title   = {Deep learning},
author  = {LeCun, Yann and Bengio, Yoshua and Hinton, Geoffrey},
journal = {Nature},
year    = {2015},
volume  = {521},
number  = {7553},
pages   = {436--444},
doi     = {10.1038/nature14539},
url     = {https://doi.org/10.1038/nature14539}
}

@article{Cawley2010,
title   = {On Over-fitting in Model Selection and Subsequent Selection Bias in Performance Evaluation},
author  = {Cawley, Gavin C. and Talbot, Nicola L. C.},
journal = {J. Mach. Learn. Res.},
year    = {2010},
volume  = {11},
number  = {70},
pages   = {2079--2107},
url     = {https://www.jmlr.org/papers/v11/cawley10a.html}
}

@article{Hawkins2004,
title   = {The Problem of Overfitting},
author  = {Hawkins, Douglas M.},
journal = {J. Chem. Inf. Comput. Sci.},
year    = {2004},
volume  = {44},
number  = {1},
pages   = {1--12},
doi     = {10.1021/ci0342472},
url     = {https://doi.org/10.1021/ci0342472}
}

@article{Butler2018,
title   = {Machine learning for molecular and materials science},
author  = {Butler, Keith T. and Davies, Daniel W. and Cartwright, Hugh and Isayev, Olexandr and Walsh, Aron},
journal = {Nature},
year    = {2018},
volume  = {559},
number  = {7715},
pages   = {547--555},
doi     = {10.1038/s41586-018-0337-2},
url     = {https://doi.org/10.1038/s41586-018-0337-2}
}

@article{Mater2019,
title   = {Deep Learning in Chemistry},
author  = {Mater, Adam C. and Coote, Michelle L.},
journal = {J. Chem. Inf. Model.},
year    = {2019},
volume  = {59},
number  = {6},
pages   = {2545--2559},
doi     = {10.1021/acs.jcim.9b00266},
url     = {https://doi.org/10.1021/acs.jcim.9b00266}
}

@article{Unke2021,
title   = {Machine Learning Force Fields},
author  = {Unke, Oliver T. and Chmiela, Stefan and Sauceda, Huziel E. and Gastegger, Michael and Poltavsky, Igor and Sch{\"u}tt, Kristof T. and Tkatchenko, Alexandre and M{\"u}ller, Klaus-Robert},
journal = {Chem. Rev.},
year    = {2021},
volume  = {121},
number  = {16},
pages   = {10142--10186},
doi     = {10.1021/acs.chemrev.0c01111},
url     = {https://doi.org/10.1021/acs.chemrev.0c01111}
}

@misc{Huang2025,
title         = {{CASCADE}: Cumulative Agentic Skill Creation through Autonomous Development and Evolution},
author        = {Huang, Xu and Chen, Junwu and Fei, Yuxing and Li, Zhuohan and Schwaller, Philippe and Ceder, Gerbrand},
year          = {2025},
eprint        = {2512.23880},
archivePrefix = {arXiv},
primaryClass  = {cs.AI},
doi           = {10.48550/arXiv.2512.23880},
url           = {https://arxiv.org/abs/2512.23880}
}

@misc{Shen2026,
title        = {An Agentic Artificial-Intelligence Framework for Automated First-Principles X-ray Photoelectron Spectroscopy Calculations on High-Performance Computing},
author       = {Shen, Hengyuan and Qian, Jin},
year         = {2026},
howpublished = {ChemRxiv},
note         = {Preprint, version 1},
doi          = {10.26434/chemrxiv.15006672/v1},
url          = {https://doi.org/10.26434/chemrxiv.15006672/v1}
}

@article{Zou2025,
title   = {{El Agente}: An autonomous agent for quantum chemistry},
author  = {Zou, Yunheng and Cheng, Austin H. and Aldossary, Abdulrahman and
         Bai, Jiaru and Leong, Shi Xuan and Campos-Gonzalez-Angulo, Jorge Arturo and
         Choi, Changhyeok and Ser, Cher Tian and Tom, Gary and Wang, Andrew and
         Zhang, Zijian and Yakavets, Ilya and Hao, Han and Crebolder, Chris and
         Bernales, Varinia and Aspuru-Guzik, Al{\'a}n},
journal = {Matter},
year    = {2025},
volume  = {8},
number  = {7},
pages   = {102263},
doi     = {10.1016/j.matt.2025.102263},
url     = {https://doi.org/10.1016/j.matt.2025.102263}
}

@article{Romera-Paredes2024,
title   = {Mathematical discoveries from program search with large language models},
author  = {Romera-Paredes, Bernardino and Barekatain, Mohammadamin and
         Novikov, Alexander and Balog, Matej and Kumar, M. Pawan and
         Dupont, Emilien and Ruiz, Francisco J. R. and Ellenberg, Jordan S. and
         Wang, Pengming and Fawzi, Omar and Kohli, Pushmeet and Fawzi, Alhussein},
journal = {Nature},
year    = {2024},
volume  = {625},
number  = {7995},
pages   = {468--475},
doi     = {10.1038/s41586-023-06924-6},
url     = {https://doi.org/10.1038/s41586-023-06924-6}
}

@misc{Novikov2025,
title         = {{AlphaEvolve}: A coding agent for scientific and algorithmic discovery},
author        = {Novikov, Alexander and V{\~u}, Ng{\^a}n and Eisenberger, Marvin and
               Dupont, Emilien and Huang, Po-Sen and Wagner, Adam Zsolt and
               Shirobokov, Sergey and Kozlovskii, Borislav and Ruiz, Francisco J. R. and
               Mehrabian, Abbas and Kumar, M. Pawan and See, Abigail and
               Chaudhuri, Swarat and Holland, George and Davies, Alex and
               Nowozin, Sebastian and Kohli, Pushmeet and Balog, Matej},
year          = {2025},
eprint        = {2506.13131},
archivePrefix = {arXiv},
primaryClass  = {cs.AI},
doi           = {10.48550/arXiv.2506.13131},
url           = {https://arxiv.org/abs/2506.13131}
}

@misc{Duston2026,
title         = {Agentic Discovery of Exchange-Correlation Density Functionals},
author        = {Duston, Titouan and Liang, Jiashu and Wang, Yuanheng and
               Gao, Weihao and Wen, Xuelan and Sheng, Nan and Ren, Weiluo and
               Sun, Yang and Chen, Yixiao},
year          = {2026},
eprint        = {2605.05460},
archivePrefix = {arXiv},
primaryClass  = {cs.AI},
doi           = {10.48550/arXiv.2605.05460},
url           = {https://arxiv.org/abs/2605.05460}
}

@misc{Bauer2026,
title         = {Autonomous Quantum Transport Measurements of {2D} Semiconductors by an {AI} Agent},
author        = {Bauer, Brandon and
                 Whalen, Matthew and
                 Watanabe, Kenji and
                 Taniguchi, Takashi and
                 Xiao, John Q. and
                 Li, Tao E. and
                 Zhao, Wenjin},
year          = {2026},
eprint        = {2609.26661},
archivePrefix = {arXiv},
primaryClass  = {cond-mat.mes-hall},
doi           = {10.48550/arXiv.2609.26661},
url           = {https://arxiv.org/abs/2609.26661}
}

@article{Weigend2005,
title   = {Balanced basis sets of split valence, triple zeta valence and quadruple zeta valence quality for {H} to {Rn}: Design and
assessment of accuracy},
author  = {Weigend, Florian and Ahlrichs, Reinhart},
journal = {Phys. Chem. Chem. Phys.},
year    = {2005},
volume  = {7},
number  = {18},
pages   = {3297--3305},
doi     = {10.1039/B508541A},
url     = {https://doi.org/10.1039/B508541A}
}

@inproceedings{Boerner2023,
address = {New York, NY, USA},
author = {Boerner, Timothy J. and Deems, Stephen and Furlani, Thomas R. and Knuth, Shelley L. and Towns, John},
booktitle = {Practice and Experience in Advanced Research Computing},
doi = {10.1145/3569951.3597559},
isbn = {9781450399852},
month = {jul},
pages = {173--176},
publisher = {ACM},
title = {{ACCESS: Advancing Innovation: NSF's Advanced Cyberinfrastructure Coordination Ecosystem: Services \& Support}},
url = {https://dl.acm.org/doi/pdf/10.1145/3569951.3597559 https://dl.acm.org/doi/10.1145/3569951.3597559},
year = {2023}
}

@article{Cui2025,
  author  = {Cui, Qiang},
  title   = {Machine learning in molecular biophysics: Protein allostery, multi-level free energy simulations, and lipid phase transitions},
  journal = {Biophysics Rev.},
  volume  = {6},
  number  = {1},
  pages   = {011305},
  year    = {2025},
  doi     = {10.1063/5.0248589}
}

@article{Gadde2025,
  author  = {Gadde, Rohit S. K. and Devaguptam, Sreelaya and
             Ren, Fangning and Mittal, Rajat and Dong, Lechen and
             Wang, Yao and Liu, Fang},
  title   = {Chatbot-assisted quantum chemistry for explicitly solvated molecules},
  journal = {Chem. Sci.},
  volume  = {16},
  number  = {9},
  pages   = {3852--3864},
  year    = {2025},
  doi     = {10.1039/D4SC08677E}
}

@article{Ju2021,
  author  = {Ju, Cheng-Wei and French, Ethan J. and Geva, Nadav
             and Kohn, Alexander W. and Lin, Zhou},
  title   = {Stacked Ensemble Machine Learning for Range-Separation Parameters},
  journal = {J. Phys. Chem. Lett.},
  volume  = {12},
  number  = {39},
  pages   = {9516--9524},
  year    = {2021},
  doi     = {10.1021/acs.jpclett.1c02506}
}

@article{Lee2026,
  author  = {Lee, Jaehyung and Ely, Justin and Zhang, Kent
             and Ajith, Akshaya and Campbell, Charles Rhys
             and Choudhary, Kamal},
  title   = {{AGAPI-Agents}: An Open-Access Agentic {AI} Platform
             for Accelerated Materials Design on {AtomGPT.org}},
  journal = {J. Phys. Chem. Lett.},
  volume  = {17},
  number  = {26},
  pages   = {7221--7231},
  year    = {2026},
  doi     = {10.1021/acs.jpclett.6c00837}
}

@article{Tran2026,
  author  = {Tran, Huan and Mahmood, Akhlak and Chaudhari, Harshal
             and Mamtani, Kuldeep and Kim, Chiho and Ramprasad, Rampi
             and Krishnamoorthy, Anand and Patra, Abhirup},
  title   = {Accelerated Design of Proton Exchange Membranes for
             Green Hydrogen Production with Artificial Intelligence},
  journal = {Chem. Mater.},
  volume  = {38},
  number  = {16},
  pages   = {8125--8134},
  year    = {2026},
  doi     = {10.1021/acs.chemmater.6c00265}
}

@article{Jablonka2023,
  author  = {Jablonka, Kevin Maik
             and Ai, Qianxiang
             and Al-Feghali, Alexander
             and Badhwar, Shruti
             and Bocarsly, Joshua D.
             and Bran, Andres M.
             and Bringuier, Stefan
             and Brinson, L. Catherine
             and Choudhary, Kamal
             and Circi, Defne
             and Cox, Sam
             and de Jong, Wibe A.
             and Evans, Matthew L.
             and Gastellu, Nicolas
             and Genzling, Jerome
             and Gil, Mar{\'i}a Victoria
             and Gupta, Ankur K.
             and Hong, Zhi
             and Imran, Alishba
             and Kruschwitz, Sabine
             and Labarre, Anne
             and L{\'a}la, Jakub
             and Liu, Tao
             and Ma, Steven
             and Majumdar, Sauradeep
             and Merz, Garrett W.
             and Moitessier, Nicolas
             and Moubarak, Elias
             and Mouri{\~n}o, Beatriz
             and Pelkie, Brenden
             and Pieler, Michael
             and Ramos, Mayk Caldas
             and Rankovi{\'c}, Bojana
             and Rodriques, Samuel G.
             and Sanders, Jacob N.
             and Schwaller, Philippe
             and Schwarting, Marcus
             and Shi, Jiale
             and Smit, Berend
             and Smith, Ben E.
             and Van Herck, Joren
             and V{\"o}lker, Christoph
             and Ward, Logan
             and Warren, Sean
             and Weiser, Benjamin
             and Zhang, Sylvester
             and Zhang, Xiaoqi
             and Zia, Ghezal Ahmad
             and Scourtas, Aristana
             and Schmidt, K. J.
             and Foster, Ian
             and White, Andrew D.
             and Blaiszik, Ben},
  title   = {14 examples of how {LLMs} can transform materials science
             and chemistry: a reflection on a large language model hackathon},
  journal = {Digital Discov.},
  volume  = {2},
  number  = {5},
  pages   = {1233--1250},
  year    = {2023},
  doi     = {10.1039/D3DD00113J}
}

@article{Adamo1999,
  author  = {Adamo, Carlo and Barone, Vincenzo},
  title   = {Toward reliable density functional methods without adjustable parameters: The {PBE0} model},
  journal = {J. Chem. Phys.},
  volume  = {110},
  pages   = {6158--6170},
  year    = {1999},
  doi     = {10.1063/1.478522}
}

\end{document}


\newpage{}

\tableofcontents{}

\newpage

\section{1. FermiLink Optimization Details}

\begin{figure}[H]
\centering
\includegraphics[width=\linewidth]{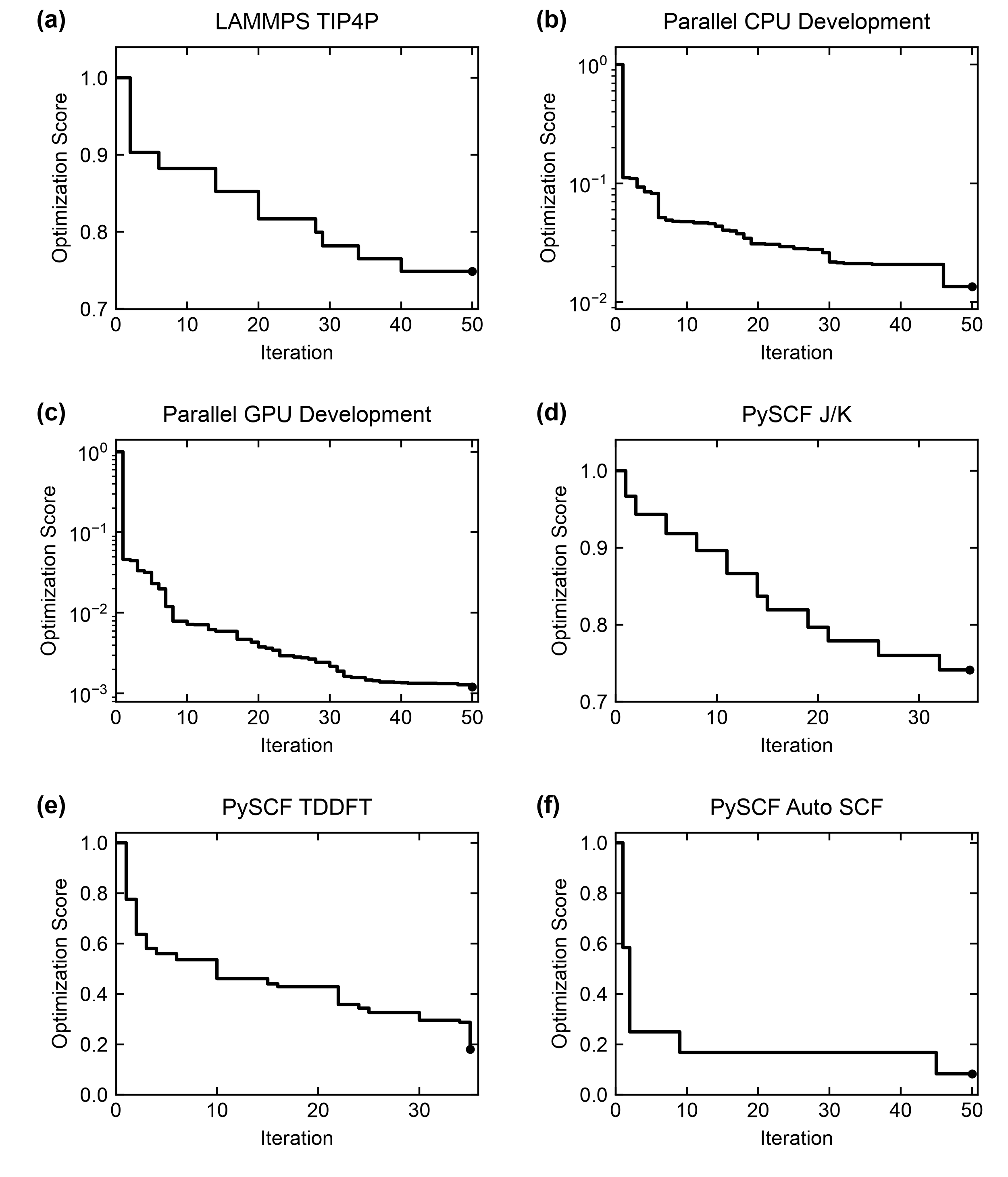}
\caption{\textsc{FermiLink} optimization trajectories for each of the targets in Table~\ref{tbl:tableS1}. (a) is the same as the black curve shown in Figs.~2a and~2b in the main manuscript. (b) and (c) are the same as the blue and green curves, respectively, in Fig.~4a of the main manuscript.}
\label{fig:fermilink-trajectories}
\end{figure}

Optimization parameters (Table~\ref{tbl:tableS1}), molecular systems used for training and testing (Tables~\ref{tbl:tableS2}--\ref{tbl:tableS6}) and completed \textsc{FermiLink} optimization trajectories (Fig.~\ref{fig:fermilink-trajectories}) are presented in this section. Other optimization input files, such as goal files and submit scripts, along with outputs are available through the Zenodo link as mentioned in the main paper. All \textsc{FermiLink} optimizations in Table~\ref{tbl:tableS1} were conducted using OpenAI's GPT-5.6-sol model with \texttt{xhigh} reasoning effort, with the exception of PySCF Auto SCF which used \texttt{max} reasoning effort, between the months of July and September 2026. Note that the GPT-5.5, GPT-5.4, and the additional GPT-5.6-sol optimizations shown in Fig.~2 in the main paper are not included in this section but can be found at the Zenodo link. The geometries in Tables~\ref{tbl:tableS4}--\ref{tbl:tableS6} were obtained from references\cite{Bolton2011,Slattery2024,QChem2025} or were constructed as model systems. The PySCF calculations in Tables~\ref{tbl:tableS4}--\ref{tbl:tableS6} used a combination of electronic basis sets and HF or DFT. These basis sets are: 3-21G, \cite{Binkley1980,Dobbs1987a,Dobbs1987b} 6-31G,\cite{Ditchfield1971,Hehre1972} 6-31G*,\cite{Ditchfield1971,Hehre1972,Hariharan1973} def2-SVP, def2-TZVP, def2-TZVPP,\cite{Weigend2005} def2-SVPD,\cite{Rappoport2010} ma-def2-TZVP,\cite{Zheng2011} def2-mSVP,\cite{Grimme2015} aug-cc-pVDZ, aug-cc-pVQZ,\cite{Dunning1989,Kendall1992,Woon1993} and aug-cc-pwCVTZ.\cite{Woon1993,Peterson2002,Balabanov2005} The DFT functionals employed include: PBE,\cite{Perdew1996} PBE0,\cite{Adamo1999} B3LYP,\cite{Becke1993,Lee1988,Stephens1994} CAM-B3LYP,\cite{Yanai2004} BP86,\cite{Becke1988,Perdew1986} M06-L,\cite{Zhao2006} r$^2$SCAN\cite{Furness2020} and PBEh-3c.\cite{Grimme2015}

\begin{table}[H]
\centering
\SITableFont
\caption{\textsc{FermiLink} Optimization Parameters for each Scientific Package and Corresponding Method}
\label{tbl:tableS1}

\setlength{\tabcolsep}{3pt}
\renewcommand{\arraystretch}{1.15}

\begin{tabular*}{\linewidth}{@{\extracolsep{\fill}}cc>{\centering\arraybackslash}m{0.17\linewidth}>{\centering\arraybackslash}m{0.13\linewidth}>{\centering\arraybackslash}
m{0.28\linewidth}@{}}
\toprule
Target & Max. & Max. Consecutive & Minimum & Simplified Meaning \\
       & Iterations & Rejections\textsuperscript{a} & Improvement\textsuperscript{b} & of Optimization Score\textsuperscript{c} \\
\midrule
LAMMPS TIP4P & 50 & 12 & 2\% & Total pair + k-space time \\
Parallel CPU Development & 50 & None & 1\% & Total MD simulation time \\
Parallel GPU Development & 50 & None & 1\% & Total MD simulation time \\
PySCF J/K & 35 & 10 & 2\% & SCF time \\
PySCF TDDFT & 35 & 10 & 2\% & TDDFT time \\
PySCF Auto SCF & 50 & None & 0\% & SCF convergence reliability and time \\
\bottomrule
\end{tabular*}

\medskip
\begin{minipage}{\linewidth}
\SITableFont
\raggedright
\textsuperscript{a}\,Optimization stops if there is a streak of rejections that meets or exceeds this threshold. Rejections occur when the code fails to meet scientific
correctness or fails to improve the previously accepted iteration's optimization score by at least the minimum improvement percentage.
\par\smallskip
\textsuperscript{b}\,The current iteration's optimization score must improve the previously accepted optimization score by at least this percentage amount in order to qualify
as an accepted code change.
\par\smallskip
\textsuperscript{c}\,Only the results of testing calculations are used to obtain an optimization score by the controller agent.
\par
\end{minipage}
\end{table}

\begin{table}[H]
\centering
\SITableFont
\caption{Training and Testing Cases Used for the
LAMMPS TIP4P Optimization\textsuperscript{a}}
\label{tbl:tableS2}

\setlength{\tabcolsep}{12pt}
\renewcommand{\arraystretch}{1.15}

\begin{tabular}{@{}cS[table-format=2]S[table-format=5,group-digits=integer,group-separator={,},group-minimum-digits=4]@{}}
\toprule
Set & \multicolumn{1}{c}{MPI Ranks} & \multicolumn{1}{c}{MD Steps} \\
\midrule
Training & 16 & 1200 \\
Training & 32 & 1200 \\
Training & 16 & 10000 \\
\midrule
Testing & 32 & 10000 \\
Testing & 64 & 1200 \\
\bottomrule
\end{tabular}

\medskip
\begin{minipage}{\linewidth}
\SITableFont
\raggedright
\textsuperscript{a}\,All cases used the same starting configuration of 13,824 flexible q-TIP4P/F water molecules simulated with periodic boundary conditions in all three spatial directions. Initial velocities were sampled from a Maxwell-Boltzmann distribution at 300 K. All were simulated in the NVE ensemble using a 0.5 fs timestep.
\par
\end{minipage}
\end{table}

\begin{table}[H]
\centering
\SITableFont
\caption{Training and Testing Cases Used for the
Parallel CPU and GPU Optimizations\textsuperscript{a}}
\label{tbl:tableS3}

\setlength{\tabcolsep}{8pt}
\renewcommand{\arraystretch}{1.15}

\begin{tabular}{@{}cS[table-format=3]S[table-format=3]cS[table-format=2]@{}}
\toprule
Set & \multicolumn{1}{c}{Water Molecules} & \multicolumn{1}{c}{Temperature / K} & Ensemble & \multicolumn{1}{c}{MD Steps} \\
\midrule
Training & 32 & 300 & NVE & 30 \\
Training & 32 & 300 & NVT & 30 \\
Training & 256 & 280 & NVE & 30 \\
Training & 256 & 280 & NVT & 30 \\
Training & 864 & 320 & NVE & 30 \\
Training & 864 & 320 & NVT & 30 \\
\midrule
Testing & 64 & 290 & NVE & 30 \\
Testing & 64 & 290 & NVT & 30 \\
Testing & 216 & 310 & NVE & 30 \\
Testing & 216 & 310 & NVT & 30 \\
Testing & 432 & 300 & NVE & 30 \\
Testing & 432 & 300 & NVT & 30 \\
\bottomrule
\end{tabular}

\medskip
\begin{minipage}{\linewidth}
\SITableFont
\raggedright
\textsuperscript{a}\,All cases used flexible q-TIP4P/F water with different numbers of water molecules. Simulations were conducted using periodic boundary conditions in all three spatial directions and a 0.5 fs timestep. Initial velocities were sampled from a Maxwell-Boltzmann distribution at the listed temperature. This temperature also served as the Langevin\cite{Schneider1978} thermostat target for NVT simulations.
\par
\end{minipage}
\end{table}

\begin{table}[H]
\centering
\SITableFont
\caption{Training and Testing Cases Used for the
PySCF J/K Optimization}
\label{tbl:tableS4}

\setlength{\tabcolsep}{3pt}
\renewcommand{\arraystretch}{1.15}

\begin{tabular*}{\linewidth}{@{\extracolsep{\fill}}cccccS[table-format=-1]S[table-format=1]@{}}
\toprule
Set & System & Method & Basis & Reference State & \multicolumn{1}{c}{Charge} & \multicolumn{1}{c}{Multiplicity} \\
\midrule
Training & Naproxen & PBE0 & def2-SVP & RKS & 0 & 1 \\
Training & Azobenzene & HF & def2-SVPD & RHF & 0 & 1 \\
Training & Pentacene & PBE & def2-SVP & RKS & 0 & 1 \\
Training & BHT Phenoxyl Radical & CAM-B3LYP & def2-SVP & ROKS & 0 & 2 \\
Training & Ubiquinone-0 Semiquinone & PBE0 & def2-SVPD & UKS & -1 & 2 \\
Training & Cyclobutane\textsuperscript{a} & HF & def2-TZVP & RHF & 0 & 1 \\
Training & \textit{n}-Dodecane\textsuperscript{a} & r$^{2}$SCAN & def2-SVP & RKS & 0 & 1 \\
\midrule
Testing & Warfarin & PBE0 & def2-SVP & RKS & 0 & 1 \\
Testing & Thalidomide & HF & def2-SVP & RHF & 0 & 1 \\
Testing & Porphyrin & PBE & def2-SVP & RKS & 0 & 1 \\
Testing & Resveratrol Phenoxyl Radical & CAM-B3LYP & def2-SVP & ROKS & 0 & 2 \\
Testing & Di-tert-butyl Nitroxide & PBE0 & def2-TZVP & UKS & 0 & 2 \\
Testing & Cyclobutane\textsuperscript{a} & HF & def2-TZVP & RHF & 0 & 1 \\
Testing & \textit{n}-Dodecane\textsuperscript{a} & r$^{2}$SCAN & def2-SVP & RKS & 0 & 1 \\
\bottomrule
\end{tabular*}

\par\vspace{3pt}
\noindent
\begin{minipage}{\linewidth}
\raggedright
\textsuperscript{a} The cyclobutane case uses in-core electron-repulsion integral storage and serves as a negative control. The rest of the cases use direct integral evaluation. The \textit{n}-Dodecane case serves as an additional control. In this control, numerical integration of the r$^{2}$SCAN exchange-correlation functional accounted for a large fraction of the total time. Both of these control cases were included in the training and testing phases to enforce correctness and to check against unacceptable slowdowns. They did not contribute to the overall performance optimization in terms of scoring. 
\end{minipage}
\end{table}

\begin{table}[H]
\centering
\SITableFont
\caption{Training and Testing Cases Used for the PySCF TDDFT Optimization }
\label{tbl:tableS5}

\setlength{\tabcolsep}{3pt}
\renewcommand{\arraystretch}{1.15}

\begin{tabular*}{\linewidth}{@{\extracolsep{\fill}}cccccS[table-format=1]S[table-format=1]@{}}
\toprule
Set & System & Method & Basis & Reference State & \multicolumn{1}{c}{Charge} & \multicolumn{1}{c}{Multiplicity} \\
\midrule
Training & Furan & PBE0 & def2-SVP & RKS & 0 & 1 \\
Training & Benzoquinone & CAM-B3LYP & def2-SVP & RKS & 0 & 1 \\
Training & Naphthalene & B3LYP & 6-31G* & RKS & 0 & 1 \\
Training & Indole & PBE0 & 6-31G* & RKS & 0 & 1 \\
Training & Benzyl Radical & PBE0 & def2-SVP & UKS & 0 & 2 \\
Training & Pyrazine\textsuperscript{a} & BP86 & def2-TZVP & RKS & 0 & 1 \\
\midrule
Testing & Thiophene & PBE0 & def2-SVP & RKS & 0 & 1 \\
Testing & Uracil & CAM-B3LYP & def2-SVP & RKS & 0 & 1 \\
Testing & Caffeine & PBE0 & 6-31G & RKS & 0 & 1 \\
Testing & Fluorene & B3LYP & 6-31G & RKS & 0 & 1 \\
Testing & Phenoxy Radical & PBE0 & def2-SVP & UKS & 0 & 2 \\
Testing & Allyl Radical\textsuperscript{a} & PBE & def2-SVP & UKS & 0 & 2 \\
\bottomrule
\end{tabular*}

\medskip
\begin{minipage}{\linewidth}
\SITableFont
\textsuperscript{a}\,These systems used the Casida formulation while the rest of the systems used the full-response TDDFT formulation.
\end{minipage}
\end{table}

\begin{table}[H]
\centering
\SITableFont
\caption{Training and Testing Cases Used for the PySCF Adaptive SCF Development}
\label{tbl:tableS6}

\setlength{\tabcolsep}{3pt}
\renewcommand{\arraystretch}{1.15}

\begin{tabular*}{\linewidth}{@{\extracolsep{\fill}}cccccS[table-format=1]S[table-format=1]@{}}
\toprule
Set & System & Method & Basis & Reference State & \multicolumn{1}{c}{Charge} & \multicolumn{1}{c}{Multiplicity} \\
\midrule
Training & CrOF & PBE & def2-TZVP & UKS & 0 & 4 \\
Training & FeCl & M06-L & ma-def2-TZVP & UKS & 0 & 6 \\
Training & Cd--Imidazole & B3LYP & 3-21G & RKS & 2 & 1 \\
Training & Cr$_2$ & B3LYP & 3-21G & RKS & 0 & 1 \\
Training & FeO & HF & def2-SVP & UHF & 0 & 5 \\
Training & NO$_3$ & HF & aug-cc-pVDZ & ROHF & 0 & 2 \\
Training & O$_3$ & HF & 6-31G* & UHF & 0 & 1 \\
Training & Ni$_2$ & HF & def2-SVP & UHF & 0 & 3 \\
Training & ZnCl\textsuperscript{a} & M06-L & ma-def2-TZVP & UKS & 0 & 2 \\
Training & S$_4$\textsuperscript{a} & PBE & ma-def2-TZVP & RKS & 0 & 1 \\
Training & MB16-43/16\textsuperscript{a} & PBEh-3c & def2-mSVP & UKS & 0 & 2 \\
Training & NF\textsuperscript{a} & HF & aug-cc-pVQZ & UHF & 0 & 3 \\
\midrule
Testing & Ferrocene & PBE0 & def2-SVP & RKS & 0 & 1 \\
Testing & DIRHIX & HF & def2-SVP & UHF & 0 & 3 \\
Testing & S$_2$ & HF & aug-cc-pVQZ & UHF & 0 & 1 \\
Testing & VCl & M06-L & ma-def2-TZVP & UKS & 0 & 5 \\
Testing & CrC & HF & def2-TZVPP & RHF & 0 & 1 \\
Testing & H$_6$ & HF & 6-31G & UHF & 0 & 1 \\
Testing & BO$_2$ & PBE & aug-cc-pVDZ & ROKS & 0 & 2 \\
Testing & FO$_2$ & PBEh-3c & def2-mSVP & UKS & 0 & 2 \\
Testing & MnS\textsuperscript{a} & M06-L & ma-def2-TZVP & UKS & 0 & 6 \\
Testing & CuCl\textsuperscript{a} & PBE & aug-cc-pwCVTZ & RKS & 0 & 1 \\
Testing & VF$_5$\textsuperscript{a} & PBE & def2-TZVP & RKS & 0 & 1 \\
Testing & ACONF/HGGG\textsuperscript{a} & PBEh-3c & def2-mSVP & RKS & 0 & 1 \\
\bottomrule
\end{tabular*}

\medskip
\begin{minipage}{\linewidth}
\SITableFont
\textsuperscript{a}\,In these well-behaved calculations, DIIS, CIAH, and OTR standalone solvers all converged to the
same final solution. These systems serve as regularization controls alongside the other systems, which demonstrate
difficult convergence behavior for one or more standalone solvers.
\end{minipage}
\end{table}

\section{2. MD Simulation Details}

The optimized LAMMPS code in Fig.~\ref{fig:fermilink-trajectories}a was assessed on two systems: a solvated DNA complex (PDB ID: 1BNA), described using the Amber OL15 force field, and a box of classical water molecules. Both systems used the rigid TIP4P-Ew water model, whereas the initial optimization and development with \textsc{FermiLink} (Fig.~\ref{fig:fermilink-trajectories}a) used the flexible q-TIP4P/F water model. Here, we describe the system preparation steps for each system, starting first with the DNA complex.

From the initial PDB structure, the crystal waters were removed and the entire complex was placed in a $6.04 \times 6.04 \times 6.04$ nm simulation box. The box was then filled with 6,616 TIP4P-Ew water molecules before counterions were added (40 Na$^+$ ions and 18 Cl$^-$ ions). Harmonic position restraints of 5 kcal mol$^{-1}$~\AA$^{-2}$ were then applied to the DNA atoms before the entire system was energy minimized using conjugate gradient. After energy minimization, the system was heated from 50 to 298.15 K in the NVT ensemble over 25 ps. After heating, the system was equilibrated in the NPT ensemble at 298.15 K and 1 atm in three 50 ps intervals with restraints applied at 5, 1, and 0.1 kcal mol$^{-1}$~\AA$^{-2}$, respectively. Afterwards, a 100 ps NPT equilibration was performed without any restraints. The bulk water system was prepared by filling a $5.07 \times 5.07 \times 5.07$ nm simulation box with 4,096 TIP4P-Ew water molecules. The system was then energy minimized with conjugate gradient prior to NVT equilibration for 25 ps at 298.15 K. Then, it was equilibrated for another 200 ps in the NPT ensemble at 298.15 K and 1 atm. Note that for all of these preparation and equilibration steps as mentioned, a 1 fs timestep was used.

After these steps, the resulting configurations and velocities were used as the initial conditions for the LAMMPS calculations in Tables~\ref{tbl:tableS7}--\ref{tbl:tableS10}. For these production simulations, we used periodic boundary conditions in all three spatial directions, along with a 10 \AA\ cutoff for Lennard-Jones and real-space electrostatic interactions, and the particle--particle particle--mesh method with a relative force accuracy of $10^{-5}$ for long-range electrostatics. A Nos\'e--Hoover thermostat at 298.15 K was used for NVT and NPT simulations, and a Nos\'e--Hoover barostat at 1 atm was used for NPT simulations. For both systems, the SHAKE algorithm was applied to water. The DNA complex simulations were performed for 500 ps using a 1 fs timestep, and the bulk water simulations were performed for 500 ps using a 2 fs timestep. The resulting trajectories from both the original and optimized LAMMPS code were found to be identical, so Tables~\ref{tbl:tableS7}--\ref{tbl:tableS10} only report the relevant timings. These timings were used to construct Figs.~2c and 2d in the main manuscript.  

\begin{table}[H]
\centering
\SITableFont
\caption{Production Simulation Wall Times for the Solvated DNA Complex Using the Original LAMMPS Code}
\label{tbl:tableS7}
\smallskip

\setlength{\tabcolsep}{3pt}
\renewcommand{\arraystretch}{1.15}

\begin{tabular*}{\linewidth}{@{\extracolsep{\fill}}cS[table-format=5.2]S[table-format=4.2]S[table-format=3.2]S[table-format=4.2]S[table-format=5.2]@{}}
\toprule
Ensemble & \multicolumn{1}{c}{Pair} & \multicolumn{1}{c}{K-space} & \multicolumn{1}{c}{Neighbor} & \multicolumn{1}{c}{Other} & \multicolumn{1}{c}{Total MD} \\
 & \multicolumn{1}{c}{Time / s} & \multicolumn{1}{c}{Time / s} & \multicolumn{1}{c}{Time / s} & \multicolumn{1}{c}{Time / s} & \multicolumn{1}{c}{Time / s} \\
\midrule
NVE & 10194.00 & 3570.30 & 758.58 & 250.12 & 14773.00 \\
NVT & 10586.00 & 3648.50 & 759.14 & 255.46 & 15249.10 \\
NPT & 14048.00 & 3575.40 & 758.89 & 1009.71 & 19392.00 \\
\bottomrule
\end{tabular*}

\medskip
\begin{minipage}{\linewidth}
\SITableFont
\end{minipage}
\end{table}

\begin{table}[H]
\centering
\SITableFont
\caption{Production Simulation Wall Times for the Solvated DNA Complex Using the Optimized LAMMPS Code}
\label{tbl:tableS8}
\smallskip

\setlength{\tabcolsep}{3pt}
\renewcommand{\arraystretch}{1.15}

\begin{tabular*}{\linewidth}{@{\extracolsep{\fill}}cS[table-format=5.2]S[table-format=4.2]S[table-format=3.2]S[table-format=4.2]S[table-format=5.2]@{}}
\toprule
Ensemble & \multicolumn{1}{c}{Pair} & \multicolumn{1}{c}{K-space} & \multicolumn{1}{c}{Neighbor} & \multicolumn{1}{c}{Other} & \multicolumn{1}{c}{Total MD} \\
 & \multicolumn{1}{c}{Time / s} & \multicolumn{1}{c}{Time / s} & \multicolumn{1}{c}{Time / s} & \multicolumn{1}{c}{Time / s} & \multicolumn{1}{c}{Time / s} \\
\midrule
NVE & 7681.80 & 2868.20 & 757.51 & 255.29 & 11562.80 \\
NVT & 8147.10 & 2977.80 & 758.38 & 266.12 & 12149.40 \\
NPT & 13337.00 & 4546.90 & 758.37 & 1013.43 & 19655.70 \\
\bottomrule
\end{tabular*}

\medskip
\begin{minipage}{\linewidth}
\SITableFont
\end{minipage}
\end{table}

\begin{table}[H]
\centering
\SITableFont
\caption{
Production Simulation Wall Times for Bulk Water Using the Original LAMMPS Code}
\label{tbl:tableS9}
\smallskip

\setlength{\tabcolsep}{3pt}
\renewcommand{\arraystretch}{1.15}

\begin{tabular*}{\linewidth}{@{\extracolsep{\fill}}cS[table-format=4.2]S[table-format=3.2]S[table-format=3.2]S[table-format=3.2]S[table-format=4.2]@{}}
\toprule
Ensemble & \multicolumn{1}{c}{Pair} & \multicolumn{1}{c}{K-space} & \multicolumn{1}{c}{Neighbor} & \multicolumn{1}{c}{Other} & \multicolumn{1}{c}{Total MD} \\
 & \multicolumn{1}{c}{Time / s} & \multicolumn{1}{c}{Time / s} & \multicolumn{1}{c}{Time / s} & \multicolumn{1}{c}{Time / s} & \multicolumn{1}{c}{Time / s} \\
\midrule
NVE & 2898.10 & 743.59 & 374.24 & 81.73 & 4097.66 \\
NVT & 2832.10 & 730.80 & 374.45 & 84.43 & 4021.78 \\
NPT & 4019.10 & 815.84 & 377.11 & 301.56 & 5513.61 \\
\bottomrule
\end{tabular*}

\medskip
\begin{minipage}{\linewidth}
\SITableFont
\end{minipage}
\end{table}

\begin{table}[H]
\centering
\SITableFont
\caption{Production Simulation Wall Times for Bulk Water Using the Optimized LAMMPS Code}
\label{tbl:tableS10}
\smallskip

\setlength{\tabcolsep}{3pt}
\renewcommand{\arraystretch}{1.15}

\begin{tabular*}{\linewidth}{@{\extracolsep{\fill}}cS[table-format=4.2]S[table-format=4.2]S[table-format=3.2]S[table-format=3.2]S[table-format=4.2]@{}}
\toprule
Ensemble & \multicolumn{1}{c}{Pair} & \multicolumn{1}{c}{K-space} & \multicolumn{1}{c}{Neighbor} & \multicolumn{1}{c}{Other} & \multicolumn{1}{c}{Total MD} \\
 & \multicolumn{1}{c}{Time / s} & \multicolumn{1}{c}{Time / s} & \multicolumn{1}{c}{Time / s} & \multicolumn{1}{c}{Time / s} & \multicolumn{1}{c}{Time / s} \\
\midrule
NVE & 2130.50 & 696.44 & 374.50 & 83.39 & 3284.83 \\
NVT & 2060.70 & 724.11 & 384.56 & 84.76 & 3254.13 \\
NPT & 3797.70 & 1172.30 & 376.67 & 303.05 & 5649.72 \\
\bottomrule
\end{tabular*}

\medskip
\begin{minipage}{\linewidth}
\SITableFont
\end{minipage}
\end{table}

The parallel CPU and GPU codes developed in Figs.~\ref{fig:fermilink-trajectories}b and~\ref{fig:fermilink-trajectories}c, respectively, were applied to systems containing 64 to 32,768 classical water molecules described using the flexible q-TIP4P/F water model. The systems were prepared by replicating 32- or 216-water configurations and adjusting the box dimensions along with molecular positions to ensure a density of 1.000~g~cm$^{-3}$. The initial velocities were then sampled at 300 K, and the simulations were performed for 2,000 steps at a 0.5 fs timestep. The resulting times (Table~\ref{tbl:tableS11}) were used to construct Fig.~4b in the main paper. 

\begin{table}[H]
\centering
\SITableFont
\caption{Production Simulation MD Times (s) for Various Water Box Sizes Using Both CPU and GPU Code\textsuperscript{a}}
\label{tbl:tableS11}
\smallskip

\setlength{\tabcolsep}{3pt}
\renewcommand{\arraystretch}{1.15}

\begin{tabular*}{\linewidth}{@{\extracolsep{\fill}}S[table-format=5,group-digits=integer,group-separator={,},group-minimum-digits=4]S[table-format=4.2]S[table-format=3.2]S[table-format=5.2]S[table-format=3.2]@{}}
\toprule
\multicolumn{1}{c}{Classical Waters} & \multicolumn{1}{c}{CPU Original} & \multicolumn{1}{c}{CPU Optimized} & \multicolumn{1}{c}{CPU Optimized} & \multicolumn{1}{c}{GPU Optimized} \\
 & \multicolumn{1}{c}{(1 Thread)} & \multicolumn{1}{c}{(1 Thread)} & \multicolumn{1}{c}{(8 Threads)} &  \\
\midrule
64 & 68.28 & 1.66 & 0.85 & 0.15 \\
216 & 301.52 & 8.40 & 2.80 & 0.25 \\
432 & 1549.75 & 24.68 & 7.65 & 0.46 \\
512 & \multicolumn{1}{c}{\textemdash} & 30.71 & 9.37 & 0.56 \\
864 & \multicolumn{1}{c}{\textemdash} & 102.77 & 23.01 & 1.37 \\
1728 & \multicolumn{1}{c}{\textemdash} & \multicolumn{1}{c}{\textemdash} & 37.24 & 2.57 \\
4096 & \multicolumn{1}{c}{\textemdash} & \multicolumn{1}{c}{\textemdash} & 136.52 & 8.98 \\
8000 & \multicolumn{1}{c}{\textemdash} & \multicolumn{1}{c}{\textemdash} & 398.35 & 22.95 \\
13824 & \multicolumn{1}{c}{\textemdash} & \multicolumn{1}{c}{\textemdash} & 1871.22 & 52.60 \\
21952 & \multicolumn{1}{c}{\textemdash} & \multicolumn{1}{c}{\textemdash} & 5461.97 & 108.79 \\
32768 & \multicolumn{1}{c}{\textemdash} & \multicolumn{1}{c}{\textemdash} & 11814.25 & 205.37 \\
\bottomrule
\end{tabular*}

\medskip
\begin{minipage}{\linewidth}
\SITableFont
\textsuperscript{a} CPU and GPU timings were obtained on an Intel Xeon Silver 4215 and an NVIDIA RTX 4090, respectively. Here, ``\textemdash'' ~indicates no simulation. 
\end{minipage}
\end{table}

\section{3. J/K Calculation Details}

The optimized code in Fig.~\ref{fig:fermilink-trajectories}d was compared against the original, unmodified code on a set of ten geometries chosen from the GMTKN55 dataset.\cite{Goerigk2017} Calculations on these systems (Table~\ref{tbl:tableS12}) used the def2-TZVP electronic basis set and were evaluated with both HF and DFT. The DFT calculations employed PBE, PBE0, B3LYP, CAM-B3LYP, and r$^2$SCAN functionals. The results of these calculations, Tables~\ref{tbl:tableS13} and~\ref{tbl:tableS14}, were used to construct Fig.~3a in the main manuscript. 

\begin{table}[H]
\centering
\SITableFont
\caption{Ten Systems Used in the PySCF J/K Post-Optimization Assessment
}
\label{tbl:tableS12}
\smallskip

\setlength{\tabcolsep}{3pt}
\renewcommand{\arraystretch}{1.15}

\begin{tabular*}{\linewidth}{@{\extracolsep{\fill}}ccccS[table-format=-1]S[table-format=1]@{}}
\toprule
System & GMTKN55 Entry & Method & Reference State & \multicolumn{1}{c}{Charge} & \multicolumn{1}{c}{Multiplicity} \\
\midrule
Radical Cation & RC21/8e & PBE0 & UKS & 1 & 2 \\
Neutral Radical & RSE43/P16 & CAM-B3LYP & ROKS & 0 & 2 \\
Sugar Conformer & SCONF/C9 & r$^2$SCAN & RKS & 0 & 1 \\
20-Water Cluster & WATER27/H2O20 & HF & RHF & 0 & 1 \\
Pericyclic Transition State & BHPERI/TS10 & HF & RHF & 0 & 1 \\
Serine Peptide & PCONF21/SER\_pII & PBE0 & RKS & 0 & 1 \\
Tryptophan Conformer & Amino20x4/TRP\_xao & CAM-B3LYP & RKS & 0 & 1 \\
Organic Isomer & ISOL24/i16e & PBE0 & RKS & 0 & 1 \\
Anthracene Dimer & IDISP/antdimer & B3LYP & RKS & 0 & 1 \\
RNA Backbone Fragment & UPU23/0a & PBE & RKS & -1 & 1 \\
\bottomrule
\end{tabular*}

\end{table}

\begin{table}[H]
\centering
\SITableFont
\caption{SCF Results for the Ten Systems in Table~\ref{tbl:tableS12} Computed Using the Original PySCF/\textsc{libcint} Code}
\label{tbl:tableS13}
\smallskip

\setlength{\tabcolsep}{3pt}
\renewcommand{\arraystretch}{1.15}

\begin{tabular*}{\linewidth}{@{\extracolsep{\fill}}cS[table-format=2]S[table-format=2]S[table-format=4.2]S[table-format=4.2]S[table-format=-4.8]@{}}
\toprule
System & \multicolumn{1}{c}{J/K} & \multicolumn{1}{c}{SCF} & \multicolumn{1}{c}{Total SCF} & \multicolumn{1}{c}{Total J/K} & \multicolumn{1}{c}{Final SCF} \\
 & \multicolumn{1}{c}{Evaluations} & \multicolumn{1}{c}{Iterations} & \multicolumn{1}{c}{Time / s} & \multicolumn{1}{c}{Time / s} & \multicolumn{1}{c}{Energy / $E_{\mathrm{h}}$} \\
\midrule
Radical Cation & 12 & 10 & 13.50 & 9.65 & -234.12329530 \\
Neutral Radical & 26 & 11 & 17.45 & 15.26 & -247.84248340 \\
Sugar Conformer & 13 & 11 & 165.18 & 140.95 & -651.16412276 \\
20-Water Cluster & 11 & 9 & 285.75 & 284.74 & -1521.40522370 \\
Pericyclic Transition State & 13 & 11 & 251.59 & 251.09 & -464.82557387 \\
Serine Peptide & 13 & 11 & 657.78 & 607.03 & -1064.93819603 \\
Tryptophan Conformer & 28 & 12 & 1150.42 & 1116.25 & -858.38230138 \\
Organic Isomer & 14 & 12 & 1836.89 & 1739.58 & -1319.19855377 \\
Anthracene Dimer & 14 & 12 & 2766.21 & 2644.40 & -1079.40504909 \\
RNA Backbone Fragment & 17 & 15 & 2884.87 & 2715.98 & -2311.39395969 \\
\bottomrule
\end{tabular*}

\medskip
\begin{minipage}{\linewidth}
\SITableFont
\end{minipage}
\end{table}

\begin{table}[H]
\centering
\SITableFont
\caption{SCF Results for the Ten Systems in Table~\ref{tbl:tableS12} Computed Using the Optimized PySCF/\textsc{libcint} Code}
\label{tbl:tableS14}
\smallskip

\setlength{\tabcolsep}{3pt}
\renewcommand{\arraystretch}{1.15}

\begin{tabular*}{\linewidth}{@{\extracolsep{\fill}}cS[table-format=2]S[table-format=2]S[table-format=4.2]S[table-format=4.2]S[table-format=-4.8]@{}}
\toprule
System & \multicolumn{1}{c}{J/K} & \multicolumn{1}{c}{SCF} & \multicolumn{1}{c}{Total SCF} & \multicolumn{1}{c}{Total J/K} & \multicolumn{1}{c}{Final SCF} \\
 & \multicolumn{1}{c}{Evaluations} & \multicolumn{1}{c}{Iterations} & \multicolumn{1}{c}{Time / s} & \multicolumn{1}{c}{Time / s} & \multicolumn{1}{c}{Energy / $E_{\mathrm{h}}$} \\
\midrule
Radical Cation & 12 & 10 & 12.79 & 8.98 & -234.12329530 \\
Neutral Radical & 26 & 11 & 14.19 & 12.01 & -247.84248340 \\
Sugar Conformer & 13 & 11 & 123.02 & 98.90 & -651.16412276 \\
20-Water Cluster & 11 & 9 & 225.53 & 224.52 & -1521.40522370 \\
Pericyclic Transition State & 13 & 11 & 185.90 & 185.40 & -464.82557387 \\
Serine Peptide & 13 & 11 & 498.99 & 448.35 & -1064.93819603 \\
Tryptophan Conformer & 28 & 12 & 858.58 & 824.53 & -858.38230138 \\
Organic Isomer & 14 & 12 & 1368.77 & 1271.76 & -1319.19855377 \\
Anthracene Dimer & 14 & 12 & 2043.42 & 1921.87 & -1079.40504909 \\
RNA Backbone Fragment & 17 & 15 & 2067.16 & 1898.81 & -2311.39395969 \\
\bottomrule
\end{tabular*}

\medskip
\begin{minipage}{\linewidth}
\SITableFont
\end{minipage}
\end{table}

\section{4. TDDFT Calculation Details}

The optimized TDDFT code in Fig.~\ref{fig:fermilink-trajectories}e was assessed on a set of ten geometries chosen from the QUESTDB dataset\cite{Veril2021,Loos2025} (Table~\ref{tbl:tableS15}) and evaluated for the first five electronic roots. All calculations used DFT with the PBE0 functional and def2-TZVP electronic basis set. The resulting Tables~\ref{tbl:tableS16} and~\ref{tbl:tableS17} indicate close agreement with the original TDDFT code but with significantly fewer Davidson iterations and Hessian on vector matrix operations. These tables were used to construct Figs.~3b--d in the main paper.

\begin{table}[H]
\centering
\SITableFont
\caption{Ten Systems Used in the PySCF TDDFT Post-Optimization Assessment}
\label{tbl:tableS15}
\smallskip

\setlength{\tabcolsep}{3pt}
\renewcommand{\arraystretch}{1.15}

\begin{tabular}{@{}ccS[table-format=1]S[table-format=1]@{}}
\toprule
System & Reference State & \multicolumn{1}{c}{Charge} & \multicolumn{1}{c}{Multiplicity} \\
\midrule
Acrolein & RKS & 0 & 1 \\
Nitromethyl Radical & UKS & 0 & 2 \\
F$_2$BS Radical & UKS & 0 & 2 \\
Octatetraene & RKS & 0 & 1 \\
Nitrobenzene & RKS & 0 & 1 \\
Adenine & RKS & 0 & 1 \\
Quinoxaline & RKS & 0 & 1 \\
Coumarin & RKS & 0 & 1 \\
Anthracene & RKS & 0 & 1 \\
BODIPY & RKS & 0 & 1 \\
\bottomrule
\end{tabular}

\end{table}

\begin{table}[H]
\centering
\SITableFont[11]
\caption{TDDFT Results for the Ten Systems in Table~\ref{tbl:tableS15} Computed Using the Original PySCF Code}
\label{tbl:tableS16}
\smallskip

\setlength{\tabcolsep}{2pt}
\renewcommand{\arraystretch}{0.65}
\setlength{\aboverulesep}{0.6pt}
\setlength{\belowrulesep}{0.6pt}

\begin{tabular*}{\linewidth}{@{\extracolsep{\fill}}cS[table-format=2]S[table-format=3]S[table-format=4.2]S[table-format=1.4]S[table-format=1.4e-1,scientific-notation=true,retain-zero-exponent=true]@{}}
\toprule
System & \multicolumn{1}{c}{Davidson} & \multicolumn{1}{c}{Response} & \multicolumn{1}{c}{TDDFT} & \multicolumn{1}{c}{Excitation} & \multicolumn{1}{c}{Oscillator} \\
 & \multicolumn{1}{c}{Iterations} & \multicolumn{1}{c}{Applications} & \multicolumn{1}{c}{Time / s} & \multicolumn{1}{c}{Energy / eV} & \multicolumn{1}{c}{Strengths} \\
\midrule
 &  &  &  & 3.6032 & 1.6471e-4 \\
 &  &  &  & 6.3696 & 3.9922e-1 \\
Acrolein & 23 & 391 & 40.23 & 6.8240 & 1.7368e-4 \\
 &  &  &  & 7.4822 & 4.6153e-4 \\
 &  &  &  & 8.0762 & 1.8859e-3 \\
\midrule
 &  &  &  & 2.0890 & 0.0000e0 \\
 &  &  &  & 2.1534 & 6.8898e-3 \\
Nitromethyl Radical & 28 & 468 & 67.34 & 2.5564 & 1.3519e-4 \\
 &  &  &  & 4.4924 & 0.0000e0 \\
 &  &  &  & 5.0495 & 6.6402e-5 \\
\midrule
 &  &  &  & 0.6309 & 0.0000e0 \\
 &  &  &  & 3.0081 & 6.9657e-4 \\
F$_2$BS Radical & 45 & 771 & 99.46 & 6.0297 & 9.6208e-4 \\
 &  &  &  & 6.0983 & 6.7622e-3 \\
 &  &  &  & 6.4639 & 1.9871e-4 \\
\midrule
 &  &  &  & 4.1170 & 1.4929e0 \\
 &  &  &  & 5.0886 & 0.0000e0 \\
Octatetraene & 25 & 428 & 599.79 & 6.1483 & 0.0000e0 \\
 &  &  &  & 6.2606 & 1.0496e-6 \\
 &  &  &  & 6.3491 & 4.8622e-4 \\
\midrule
 &  &  &  & 3.8949 & 0.0000e0 \\
 &  &  &  & 4.3574 & 2.0594e-4 \\
Nitrobenzene & 24 & 414 & 613.69 & 4.5557 & 1.5267e-2 \\
 &  &  &  & 4.9443 & 2.0845e-1 \\
 &  &  &  & 6.0720 & 9.5631e-4 \\
\midrule
 &  &  &  & 5.1430 & 3.2274e-4 \\
 &  &  &  & 5.1965 & 2.0595e-1 \\
Adenine & 27 & 451 & 1029.23 & 5.4329 & 4.2042e-2 \\
 &  &  &  & 5.7636 & 8.6153e-4 \\
 &  &  &  & 6.0856 & 3.4951e-4 \\
\midrule
 &  &  &  & 3.5786 & 2.0669e-3 \\
 &  &  &  & 4.2391 & 2.6338e-2 \\
Quinoxaline & 25 & 432 & 1086.80 & 4.5253 & 8.2048e-2 \\
 &  &  &  & 4.9301 & 0.0000e0 \\
 &  &  &  & 5.3019 & 0.0000e0 \\
\midrule
 &  &  &  & 4.2842 & 1.2670e-1 \\
 &  &  &  & 4.5598 & 9.7455e-7 \\
Coumarin & 24 & 429 & 1368.11 & 4.7646 & 1.5493e-1 \\
 &  &  &  & 5.5693 & 3.1583e-2 \\
 &  &  &  & 5.9845 & 1.5513e-1 \\
\midrule
 &  &  &  & 3.2925 & 5.5515e-2 \\
 &  &  &  & 3.9649 & 2.0785e-4 \\
Anthracene & 23 & 397 & 2923.81 & 4.7122 & 0.0000e0 \\
 &  &  &  & 4.9662 & 0.0000e0 \\
 &  &  &  & 5.3003 & 2.0185e0 \\
\midrule
 &  &  &  & 3.1639 & 4.2926e-1 \\
 &  &  &  & 3.7470 & 1.7026e-1 \\
BODIPY & 29 & 509 & 3369.15 & 3.9090 & 2.7262e-2 \\
 &  &  &  & 5.3727 & 4.0355e-2 \\
 &  &  &  & 5.4458 & 0.0000e0 \\
\bottomrule
\end{tabular*}

\smallskip
\begin{minipage}{\linewidth}
\SITableFont[11]
\end{minipage}
\end{table}

\begin{table}[H]
\centering
\SITableFont[11]
\caption{TDDFT Results for the Ten Systems in Table~\ref{tbl:tableS15} Computed Using the Optimized PySCF Code}
\label{tbl:tableS17}
\smallskip

\setlength{\tabcolsep}{2pt}
\renewcommand{\arraystretch}{0.65}
\setlength{\aboverulesep}{0.6pt}
\setlength{\belowrulesep}{0.6pt}

\begin{tabular*}{\linewidth}{@{\extracolsep{\fill}}cS[table-format=1]S[table-format=2]S[table-format=3.2]S[table-format=1.4]S[table-format=1.4e-1,scientific-notation=true,retain-zero-exponent=true]@{}}
\toprule
System & \multicolumn{1}{c}{Davidson} & \multicolumn{1}{c}{Response} & \multicolumn{1}{c}{TDDFT} & \multicolumn{1}{c}{Excitation} & \multicolumn{1}{c}{Oscillator} \\
 & \multicolumn{1}{c}{Iterations} & \multicolumn{1}{c}{Applications} & \multicolumn{1}{c}{Time / s} & \multicolumn{1}{c}{Energy / eV} & \multicolumn{1}{c}{Strengths} \\
\midrule
 &  &  &  & 3.6032 & 1.6471e-4 \\
 &  &  &  & 6.3696 & 3.9922e-1 \\
Acrolein & 8 & 47 & 4.91 & 6.8240 & 1.7367e-4 \\
 &  &  &  & 7.4822 & 4.6153e-4 \\
 &  &  &  & 8.0762 & 1.8859e-3 \\
\midrule
 &  &  &  & 2.0890 & 0.0000e0 \\
 &  &  &  & 2.1534 & 6.8898e-3 \\
Nitromethyl Radical & 8 & 49 & 6.87 & 2.5564 & 1.3520e-4 \\
 &  &  &  & 4.4924 & 0.0000e0 \\
 &  &  &  & 5.0495 & 6.6394e-5 \\
\midrule
 &  &  &  & 0.6309 & 0.0000e0 \\
 &  &  &  & 3.0081 & 6.9659e-4 \\
F$_2$BS Radical & 9 & 51 & 6.44 & 6.0297 & 9.6215e-4 \\
 &  &  &  & 6.0983 & 6.7623e-3 \\
 &  &  &  & 6.4639 & 1.9874e-4 \\
\midrule
 &  &  &  & 4.1170 & 1.4929e0 \\
 &  &  &  & 5.0886 & 0.0000e0 \\
Octatetraene & 9 & 55 & 91.06 & 6.1483 & 0.0000e0 \\
 &  &  &  & 6.2606 & 1.0551e-6 \\
 &  &  &  & 6.3491 & 4.8628e-4 \\
\midrule
 &  &  &  & 3.8949 & 0.0000e0 \\
 &  &  &  & 4.3574 & 2.0593e-4 \\
Nitrobenzene & 9 & 49 & 92.64 & 4.5557 & 1.5266e-2 \\
 &  &  &  & 4.9443 & 2.0845e-1 \\
 &  &  &  & 6.0720 & 9.5629e-4 \\
\midrule
 &  &  &  & 5.1430 & 3.2274e-4 \\
 &  &  &  & 5.1965 & 2.0596e-1 \\
Adenine & 9 & 58 & 161.29 & 5.4329 & 4.2040e-2 \\
 &  &  &  & 5.7636 & 8.6159e-4 \\
 &  &  &  & 6.0856 & 3.4950e-4 \\
\midrule
 &  &  &  & 3.5786 & 2.0669e-3 \\
 &  &  &  & 4.2391 & 2.6336e-2 \\
Quinoxaline & 8 & 49 & 153.37 & 4.5253 & 8.2045e-2 \\
 &  &  &  & 4.9301 & 0.0000e0 \\
 &  &  &  & 5.3019 & 0.0000e0 \\
\midrule
 &  &  &  & 4.2842 & 1.2671e-1 \\
 &  &  &  & 4.5598 & 9.7616e-7 \\
Coumarin & 9 & 56 & 222.10 & 4.7646 & 1.5494e-1 \\
 &  &  &  & 5.5693 & 3.1583e-2 \\
 &  &  &  & 5.9845 & 1.5513e-1 \\
\midrule
 &  &  &  & 3.2925 & 5.5515e-2 \\
 &  &  &  & 3.9649 & 2.0782e-4 \\
Anthracene & 8 & 47 & 419.89 & 4.7122 & 0.0000e0 \\
 &  &  &  & 4.9662 & 0.0000e0 \\
 &  &  &  & 5.3003 & 2.0185e0 \\
\midrule
 &  &  &  & 3.1639 & 4.2926e-1 \\
 &  &  &  & 3.7470 & 1.7026e-1 \\
BODIPY & 9 & 54 & 442.54 & 3.9090 & 2.7262e-2 \\
 &  &  &  & 5.3727 & 4.0355e-2 \\
 &  &  &  & 5.4458 & 0.0000e0 \\
\bottomrule
\end{tabular*}

\smallskip
\begin{minipage}{\linewidth}
\SITableFont[11]
\end{minipage}
\end{table}

\section{5. Adaptive SCF Calculation Details}

The adaptive SCF code developed in Fig.~\ref{fig:fermilink-trajectories}f, which is denoted as AUTO(FL) in the main paper, was assessed on four systems (Table~\ref{tbl:tableS18}). The serine peptide geometry was obtained from the GMTKN55 dataset, the Fe(CO)$_4$(PH$_3$) geometry was obtained from Sethio et al.,\cite{Sethio2024} the Fe(NHC-Me)(CO)(NO)$_2$ geometry was obtained from the MTM dataset,\cite{Qin2024,FdezGalvan2025} and the [NiCl$_4$]$^{2-}$ geometry was obtained from the TM-GSspin dataset.\cite{Cho2024} The code was compared against the standalone DIIS, CIAH, and OTR solvers. Another adaptive SCF implementation, denoted as AUTO(C) in the main paper, was also assessed. All calculations used the def2-SVP electronic basis set and either HF or DFT with the PBE0 functional. These results (Tables~\ref{tbl:tableS19},~\ref{tbl:tableS20}, and~\ref{tbl:tableS21}) were used to construct Fig.~5 in the main paper. 

\begin{table}[H]
\centering
\SITableFont
\caption{Four Systems Used in the PySCF Adaptive SCF Post-Development Assessment}
\label{tbl:tableS18}
\smallskip

\setlength{\tabcolsep}{3pt}
\renewcommand{\arraystretch}{1.15}

\begin{tabular*}{\linewidth}{@{\extracolsep{\fill}}cccS[table-format=-1]S[table-format=1]@{}}
\toprule
System & Method & Reference State & \multicolumn{1}{c}{Charge} & \multicolumn{1}{c}{Multiplicity} \\
\midrule
{Serine Peptide} & PBE0 & RKS & 0 & 1 \\
{Fe(NHC-Me)(CO)(NO)$_2$} & HF & RHF & 0 & 1 \\
{Fe(CO)$_4$(PH$_3$)}\textsuperscript{a} & HF & UHF & 0 & 3 \\
{[NiCl$_4$]$^{2-}$} & PBE0 & UKS & -2 & 3 \\
\bottomrule
\end{tabular*}

\medskip
\begin{minipage}{\linewidth}
\SITableFont
\noindent\textsuperscript{a}\,This system was also included as part of the testing dataset, which is labeled as DIRHIX in Table~\ref{tbl:tableS6}.
\end{minipage}
\end{table}

\begin{table}[H]
\centering
\SITableFont
\caption{Final SCF Energies for the Four Systems in Table~\ref{tbl:tableS18}\textsuperscript{a}}
\label{tbl:tableS19}
\smallskip

\setlength{\tabcolsep}{0.75pt}
\renewcommand{\arraystretch}{1.20}

\begin{tabular*}{\linewidth}{@{\extracolsep{\fill}}cS[table-format=-4.8]S[table-format=-4.8]S[table-format=-4.8]S[table-format=-4.8]S[table-format=-4.8]@{}}
\toprule
System & \multicolumn{1}{c}{DIIS} & \multicolumn{1}{c}{CIAH} & \multicolumn{1}{c}{OTR} & \multicolumn{1}{c}{AUTO(FL)} & \multicolumn{1}{c}{AUTO(C)} \\
 & \multicolumn{1}{c}{Energy / $E_{\mathrm{h}}$} & \multicolumn{1}{c}{Energy / $E_{\mathrm{h}}$} & \multicolumn{1}{c}{Energy / $E_{\mathrm{h}}$} & \multicolumn{1}{c}{Energy / $E_{\mathrm{h}}$} & \multicolumn{1}{c}{Energy / $E_{\mathrm{h}}$} \\
\midrule
{Serine Peptide} & -1063.75995716 & -1063.75995716 & -1063.75995716 & -1063.75995716 & -1063.75995716 \\
{\ensuremath{\vcenter{\hbox{\shortstack[c]{Fe(NHC-Me)\\(CO)(NO)$_2$}}}}} & \multicolumn{1}{c}{x} & -1935.44391723 & -1935.44663408 & -1935.44663408 & -1935.44663408 \\
{Fe(CO)$_4$(PH$_3$)} & \multicolumn{1}{c}{x} & -2054.92825933 & -2054.96719912 & -2054.96721217 & -2054.96721215 \\
{[NiCl$_4$]$^{2-}$} & \multicolumn{1}{c}{*} & -3347.46821713 & -3347.69690475 & -3347.69690282 & -3347.69689726 \\
\bottomrule
\end{tabular*}

\medskip
\begin{minipage}{\linewidth}
\SITableFont
\textsuperscript{a}: The symbols x and * represent no convergence and convergence to an internally unstable solution, respectively.
\end{minipage}
\end{table}

\begin{table}[H]
\centering
\SITableFont
\caption{Spin Expectation Values for the Two Unrestricted Systems in Table~\ref{tbl:tableS18}\textsuperscript{a}}
\label{tbl:tableS20}
\smallskip

\setlength{\tabcolsep}{3pt}
\renewcommand{\arraystretch}{1.20}

\begin{tabular*}{\linewidth}{@{\extracolsep{\fill}}ccS[table-format=1.3]S[table-format=1.3]S[table-format=1.3]S[table-format=1.3]@{}}
\toprule
System & \multicolumn{1}{c}{DIIS} & \multicolumn{1}{c}{CIAH} & \multicolumn{1}{c}{OTR} & \multicolumn{1}{c}{AUTO(FL)} & \multicolumn{1}{c}{AUTO(C)} \\
 & \multicolumn{1}{c}{$\langle S^2\rangle$} & \multicolumn{1}{c}{$\langle S^2\rangle$} & \multicolumn{1}{c}{$\langle S^2\rangle$} & \multicolumn{1}{c}{$\langle S^2\rangle$} & \multicolumn{1}{c}{$\langle S^2\rangle$} \\
\midrule
{Fe(CO)$_4$(PH$_3$)} & \multicolumn{1}{c}{x} & 2.372 & 3.326 & 3.326 & 3.326 \\
{[NiCl$_4$]$^{2-}$} & \multicolumn{1}{c}{*} & 3.005 & 2.002 & 2.002 & 2.002 \\
\bottomrule
\end{tabular*}

\medskip
\begin{minipage}{\linewidth}
\SITableFont
\textsuperscript{a}: The symbols x and * represent no convergence and convergence to an internally unstable solution, respectively.
\end{minipage}
\end{table}

\begin{table}[H]
\centering
\SITableFont
\caption{SCF Times for the Calculations in Table~\ref{tbl:tableS19}\textsuperscript{a}}
\label{tbl:tableS21}
\smallskip

\setlength{\tabcolsep}{3pt}
\renewcommand{\arraystretch}{1.20}

\begin{tabular*}{\linewidth}{@{\extracolsep{\fill}}cS[table-format=2.2]S[table-format=3.2]S[table-format=3.2]S[table-format=3.2]S[table-format=3.2]@{}}
\toprule
System & \multicolumn{1}{c}{DIIS} & \multicolumn{1}{c}{CIAH} & \multicolumn{1}{c}{OTR} & \multicolumn{1}{c}{AUTO(FL)} & \multicolumn{1}{c}{AUTO(C)} \\
 & \multicolumn{1}{c}{Time / s} & \multicolumn{1}{c}{Time / s} & \multicolumn{1}{c}{Time / s} & \multicolumn{1}{c}{Time / s} & \multicolumn{1}{c}{Time / s} \\
\midrule
{Serine Peptide} & 92.32 & 226.21 & 484.98 & 92.32 & 514.61 \\
{Fe(NHC-Me)(CO)(NO)$_2$} & \multicolumn{1}{c}{x} & 44.81 & 77.68 & 104.65 & 123.24 \\
{Fe(CO)$_4$(PH$_3$)} & \multicolumn{1}{c}{x} & 20.84 & 57.96 & 64.06 & 48.07 \\
{[NiCl$_4$]$^{2-}$} & \multicolumn{1}{c}{*} & 24.23 & 31.36 & 27.75 & 17.95 \\
\bottomrule
\end{tabular*}

\medskip
\begin{minipage}{\linewidth}
\SITableFont
\textsuperscript{a}: The symbols x and * represent no convergence and convergence to an internally unstable solution, respectively. These times do not include the post-convergence internal stability analysis, unless it is part of the adaptive SCF algorithm itself. 
\end{minipage}
\end{table}

\newpage

\begingroup
\singlespacing
\setlength{\bibsep}{0pt}

\endgroup